\documentclass[twocolumn]{aastex63}

\received{}
\revised{}
\accepted{}
\submitjournal{PASP}

\shorttitle{Apodizers in KPIC}
\shortauthors{Calvin et al.}

\graphicspath{{./}{figures/}{AP_figs/}}

\begin{document}

\title{Enhancing direct exoplanet spectroscopy with apodizing and beam shaping optics}

\correspondingauthor{Benjamin Calvin}
\email{bcalvin@astro.ucla.edu, ben.a.calv@gmail.com} 

\author[0000-0003-4737-5486]{Benjamin Calvin} 
\affiliation{California Institute of Technology,
1200 E. California Blvd.,
Pasadena, CA 91125, USA}

\author[0000-0001-5213-6207]{Nemanja Jovanovic}
\affiliation{California Institute of Technology,
1200 E. California Blvd.,
Pasadena, CA 91125, USA}

\author[0000-0003-4769-1665]{Garreth Ruane}
\affiliation{California Institute of Technology,
1200 E. California Blvd.,
Pasadena, CA 91125, USA}
\affiliation{Jet Propulsion Laboratory, 
California Institute of Technology,
4800 Oak Grove Drive, Pasadena, 
CA 91109, USA}

\author{Jacklyn Pezzato} 
\affiliation{California Institute of Technology, 
1200 E. California Blvd.,
Pasadena, CA 91125, USA}

\author{Jennah Colborn} 
\affiliation{California Institute of Technology,
1200 E. California Blvd.,
Pasadena, CA 91125, USA}

\author{Daniel Echeverri} 
\affiliation{California Institute of Technology,
1200 E. California Blvd.,
Pasadena, CA 91125, USA}

\author{Tobias Schofield} 
\affiliation{California Institute of Technology,
1200 E. California Blvd.,
Pasadena, CA 91125, USA}

\author{Michael Porter} 
\affiliation{California Institute of Technology,
1200 E. California Blvd.,
Pasadena, CA 91125, USA}

\author{J. Kent Wallace}
\affiliation{Jet Propulsion Laboratory, 
California Institute of Technology,
4800 Oak Grove Drive, Pasadena, 
CA 91109, USA}

\author[0000-0001-8953-1008]{Jacques-Robert Delorme} 
\affiliation{California Institute of Technology,
1200 E. California Blvd.,
Pasadena, CA 91125, USA}
\affiliation{W. M. Keck Observatory, Kamuela, HI 96743, USA}

\author{Dimitri Mawet} 
\affiliation{California Institute of Technology,
1200 E. California Blvd.,
Pasadena, CA 91125, USA}
\affiliation{Jet Propulsion Laboratory, 
California Institute of Technology,
4800 Oak Grove Drive, Pasadena, 
CA 91109, USA}

\begin{abstract}
Direct exoplanet spectroscopy aims to measure the spectrum of an exoplanet while simultaneously minimizing the light collected from its host star. Isolating the planet light from the starlight improves the signal-to-noise ratio (S/N) per spectral channel when noise due to the star dominates, which may enable new studies of the exoplanet atmosphere with unprecedented detail at high spectral resolution ($>$30,000). However, the optimal instrument design depends on the flux level from the planet and star compared to the noise due to other sources, such as detector noise and thermal background. Here we present the design, fabrication, and laboratory demonstration of specially-designed optics to improve the S/N in two potential regimes in direct exoplanet spectroscopy with adaptive optics instruments. The first is a pair of beam-shaping lenses that increase the planet signal by improving the coupling efficiency into a single-mode fiber at the known position of the planet. The second is a grayscale apodizer that reduces the diffracted starlight for planets at small angular separations from their host star. The former especially increases S/N when dominated by detector noise or thermal background, while the latter helps reduce stellar noise. We show good agreement between the theoretical and experimental point spread functions in each case and predict the exposure time reduction ($\sim33\%$) that each set of optics provides in simulated observations of 51 Eridani b using the Keck Planet Imager and Characterizer instrument at W.M. Keck Observatory.

\end{abstract}

\keywords{astronomical instrumentation: high angular resolution, techniques: high angular resolution, exoplanets}

\section{Introduction} \label{sec:intro}

With over 4000 exoplanets confirmed to date, detection has given way to the era of characterization, critical to understanding the properties of these systems. In this vein, high contrast imagers (HCI), which isolate the light from the planet, offer numerous advantages over indirect techniques that rely on the signal from the host star alone. Typically, HCI use advanced wavefront control techniques combined with coronagraphs to extinguish the star light and minimize contamination to the planet signal~\citep{macintosh2014GPI,jovanovic2015SCE,males2018MAG,beuzit2019SPH}. These systems often exploit low to medium resolving power (R$\sim$10-1,000), integral field spectrographs to characterize the targets (e.g. the CHARIS instrument~\citep{groff2016LTP}). Besides providing spectral information about the target, dispersing the light offers two advantages. Firstly, the distance between a speckle and the optical axis increases linearly with wavelength, while a planet remains a fixed distance away, which can be used to differentiate between the two. Indeed, this technique is referred to as spectral differential imaging (SDI)\citep{Sparks2002}. Secondly, the spectrum of a planet has a different profile to that of a star, which enables the observer to constrain the nature of the orbiting object. This technique has been exploited to detect or characterize both planets~\citep{barman2015SDW} as well as disks~\citep{currie2017SSF}. 

Recently, the field has shifted focus towards much greater resolving power (R$>$30,000), to really take advantage of SDI in the regime where spectral lines in the target begin to be resolved~\citep{snellen2015CHD}. The technique, dubbed \textit{High dispersion Coronagraphy (HDC)}, optimally combines high contrast imaging techniques such as adaptive optics/wavefront control plus coronagraphy to high resolution spectroscopy~\citep{wang2017OEH,mawet2017OEH}. One approach that has been explored is to use an optical fiber to route the light of the known planet from the focal planet to the spectrograph. A single-mode fiber (SMF) is the ideal transport vehicle owing to the fact it has a field-of-view (FOV) that can be matched to the 1~$\lambda/D$ width of the point spread function (PSF), enabling efficient coupling for the planet light and suppression of unwanted star light. Its small core size only allows light to be guided in a single mode providing spatial filtering, which further suppresses starlight speckles from coupling in and more importantly delivers an ultra stable output beam profile, highly desired by many instruments~\citep{schwab2012-SME,woillez2003-IIS,crass2020-FDO}. In addition, its narrow FOV reduces the amount of sky background injected into the spectrograph, which can be several orders of magnitude greater in the case of a seeing-limited spectrograph. 

HDC provides the ability to do species-by-species molecular characterization (e.g.~oxygen, water, carbon dioxide, methane), thermal (vertical) atmospheric structure, planetary spin measurements (length of day), and potentially Doppler imaging of atmospheric (clouds) and/or surface features (continents versus oceans)~\citep{wang2017OEH}. As such several projects have been initiated to realize this new technique from the ground including: the Keck Planet Imager and Characterizer (KPIC)~\citep{jovanovic2019-KPI}, which combines Keck AO and NIRSPEC, the Rigorous Exoplanetary Atmosphere Characterization with High dispersion coronography instrument (REACH)~\citep{jovanovic2017DPC}, which combines SCExAO and IRD and High-Resolution Imaging and Spectroscopy of Exoplanets (HiRISE), which combines SPHERE and CRIRES+\citep{vigan2018BHS}. The phase I version of KPIC and the REACH instrument are both transitioning from commissioning to early science at the time of writing of this article and offer complimentary wavelength coverage across the near-IR (NIR) on Maunakea (REACH operates from y-H and KPIC operates in K and L bands), while HiRISE is still in the development stage. 

Key to being able to disperse the faint planet signal across many pixels and maintain signal-to-noise (S/N), as is the case in a high resolution spectrograph, is maximizing the planet throughput. This can be done by improving the adaptive optics (AO) correction to boost the Strehl ratio and hence coupling efficiency, reshaping the PSF to better match the fiber mode~\citep{jovanovic2017EIL}, or minimizing the number of optics in the system. In addition, it is also important to minimize the stellar leakage into the planet fiber to reduce the photon noise contribution from the star. Improved AO correction can also help here, but specific focal plane wavefront control techniques such as speckle nulling~\citep{bottom2016SNW,Sayson2019} or the use of a pupil plane apodizing mask~\citep{zhang2018CMA}, are better suited to reducing speckle noise. Although these technologies achieve different goals, they are used to reduce total exposure time, which is critical when characterizing faint exoplanets with high spectral resolution on large telescopes.

The technology or approach of choice depends on the parameters of the system being observed, such as planet-to-star flux ratio, angular separation, and the waveband of observation. In this paper, we present the development of two technologies that can help boost the performance of HDC on KPIC and similar instruments, namely: Phase Induced Amplitude Apodization (PIAA) optics for re-shaping the beam and boosting coupling and a grayscale microdot apodizer (MDA) for suppressing diffraction features at small angular separations. We aim to deploy these as part of the phase II upgrade of the KPIC instrument in late 2021. In Section~\ref{sec:tech} we outline the two technologies and present the design and simulated performance of such devices. In Section~\ref{sec:methods} we present the experimental setup and procedures to evaluate the two technologies and in Section~\ref{sec:results} we summarize and compare the results to simulations. Section~\ref{sec:discussion} compares the two technologies in the context of a hypothetical observing scenario to highlight the benefits of each in their respective domains and Section~\ref{sec:conclusions} rounds out the paper with some concluding thoughts.

\section{Apodization Optics} \label{sec:tech}

\begin{figure}[t]
\centering
\includegraphics[width = \linewidth]{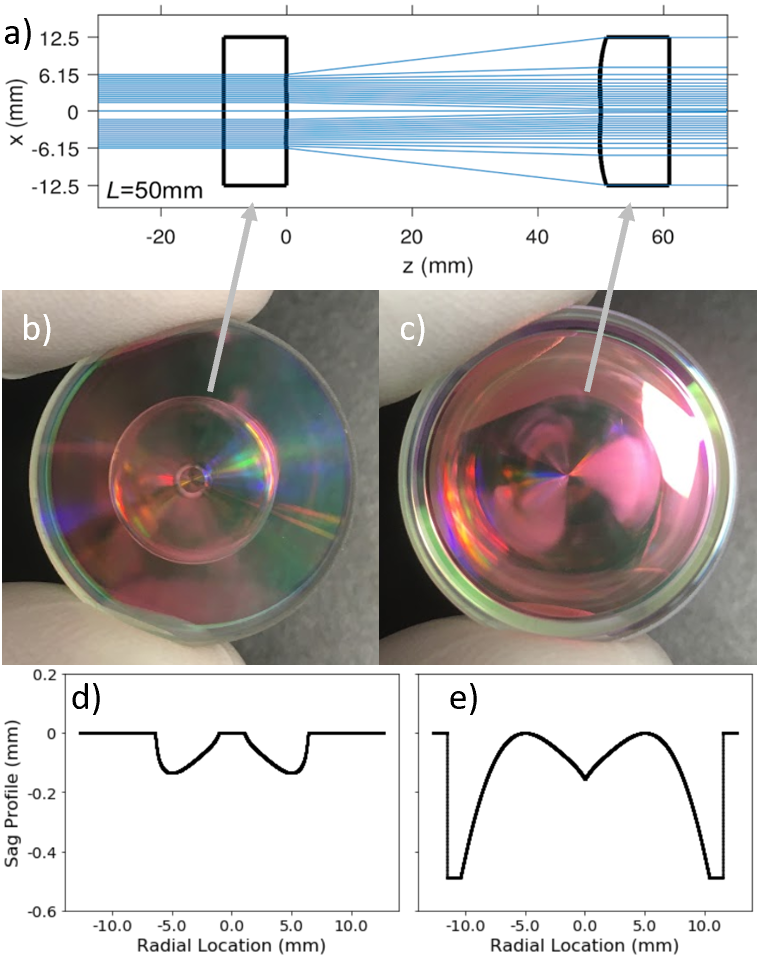}
\caption{a) A ray trace through the PIAA optics. Here, the 50 mm represents the inner spacing between the powered surfaces of the PIAA optics. b) An image of the first lens, which refracts the beam to reshape it. c) An image of the second lens in the pair, which re-collimates the light after beam shaping. d) The radial sag profile for the first lens. e) The radial sag profile for the second lens. Both sag profiles are azimuthally invariant.\label{fig:piaa}}
\end{figure}

\begin{figure}[t]
\centering
\includegraphics[width = \linewidth]{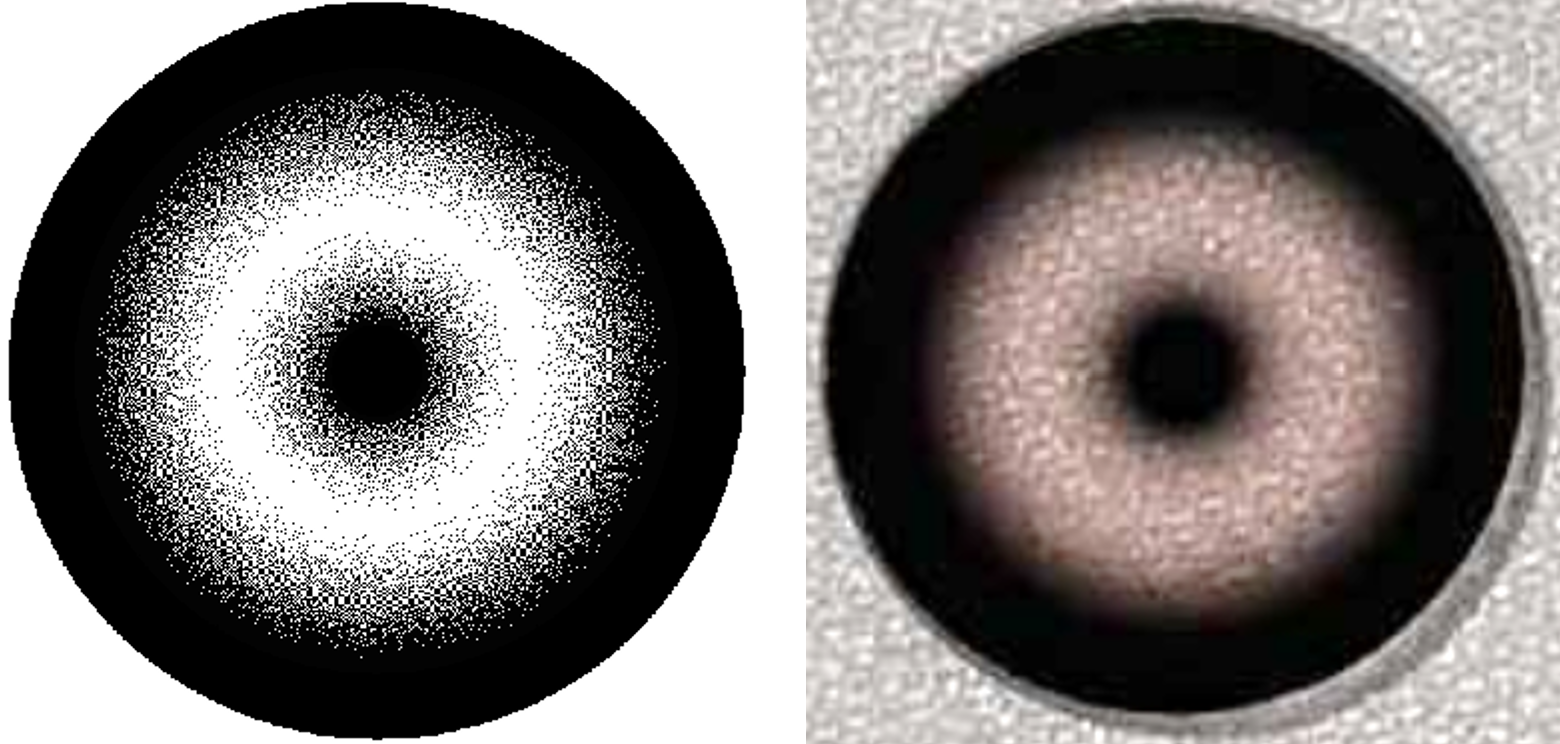}
\caption{(Left) Design for the MDA. It consists of a 500$\times$500 grid of 25 $\mu$m chrome squares forming a 12.3 mm annulus of transmission. The white areas show the region of high transmission, while the black parts are covered with chrome. For scale, the outer black edge of the MDA is 15 mm. (Right) A white-light photograph of a device manufactured for KPIC placed on textured background showing the transmission function in the visible regime appears similar to the design. \label{fig:apod}}
\end{figure}

This section provides an overview of the two technologies that we have developed to enhance the S/N of direct spectroscopic, or HDC, observations of exoplanets: PIAA (see Fig.~\ref{fig:piaa}) and MDA (see Fig.~\ref{fig:apod}) optics. Each has been employed previously in apodized coronagraphs in the context of high-contrast imaging \citep{Watson1991,Nisenson2001,Kasdin2003,Soummer2003_APLC}, where the PIAA \citep{guyon2003PIA} or MDA \citep{Dorrer2007,Martinez2009a} modifies the shape of the beam at, or near, the re-imaged telescope pupil to reduce diffracted starlight at small angular separations where exoplanets may be directly imaged. These pupil-plane optics do not require any focal plane masks to work for our purposes with KPIC. Our application differs from conventional high contrast imaging in that we aim to maximize the S/N in the measured spectrum of an exoplanet using a diffraction-limited spectrograph that is fed by a SMF. As such, the PIAA and MDA optics that we developed for KPIC are optimized to maximize the coupling efficiency for planet light, $\eta_p$, and minimize the fraction of the starlight that is coupled into the SMF, $\eta_s$.

The coupling efficiency of a coherent, scalar field, $E(\mathbf{r})$, into a fiber mode $\Psi(\mathbf{r})$ is given by:
\begin{equation}
    \eta =\frac{\left|\int E^*(\mathbf{r}) \Psi(\mathbf{r})dA\right|^2}{\int \left| E(\mathbf{r})\right|^2dA \int \left|\Psi(\mathbf{r})\right|^2dA}, 
    \label{eqn:couplingeff}
\end{equation}
where $\mathbf{r}$ is the coordinate in the plane transverse to the beam at the fiber. We use Eqn.~\ref{eqn:couplingeff}, often referred to as the overlap integral, to compute $\eta_p$ and $\eta_s$ by plugging in the field at the fiber, $E(\mathbf{r})$, due to the planet and star, respectively. When stellar photon noise dominates, $S/N \propto \eta_p/\sqrt{\eta_s}$, while in most other cases $S/N \propto \eta_p$. In essence, the PIAA optics are designed for the latter scenario, whereas the MDA is optimized for the former. 

The PIAA optics consist of a pair of beam shaping lenses designed to alter the effective distribution of light at the pupil, in order to reshape the PSF in the focal plane. In the original implementation, PIAA lenses were used to reduce the Airy rings in coronagraphs~\citep{guyon2003PIA}. Later, PIAA optics were used to optimize the coupling to a SMF~\citep{jovanovic2017EIL}.  
\begin{figure*}[t]
\centering
\includegraphics[width = 0.8\textwidth]{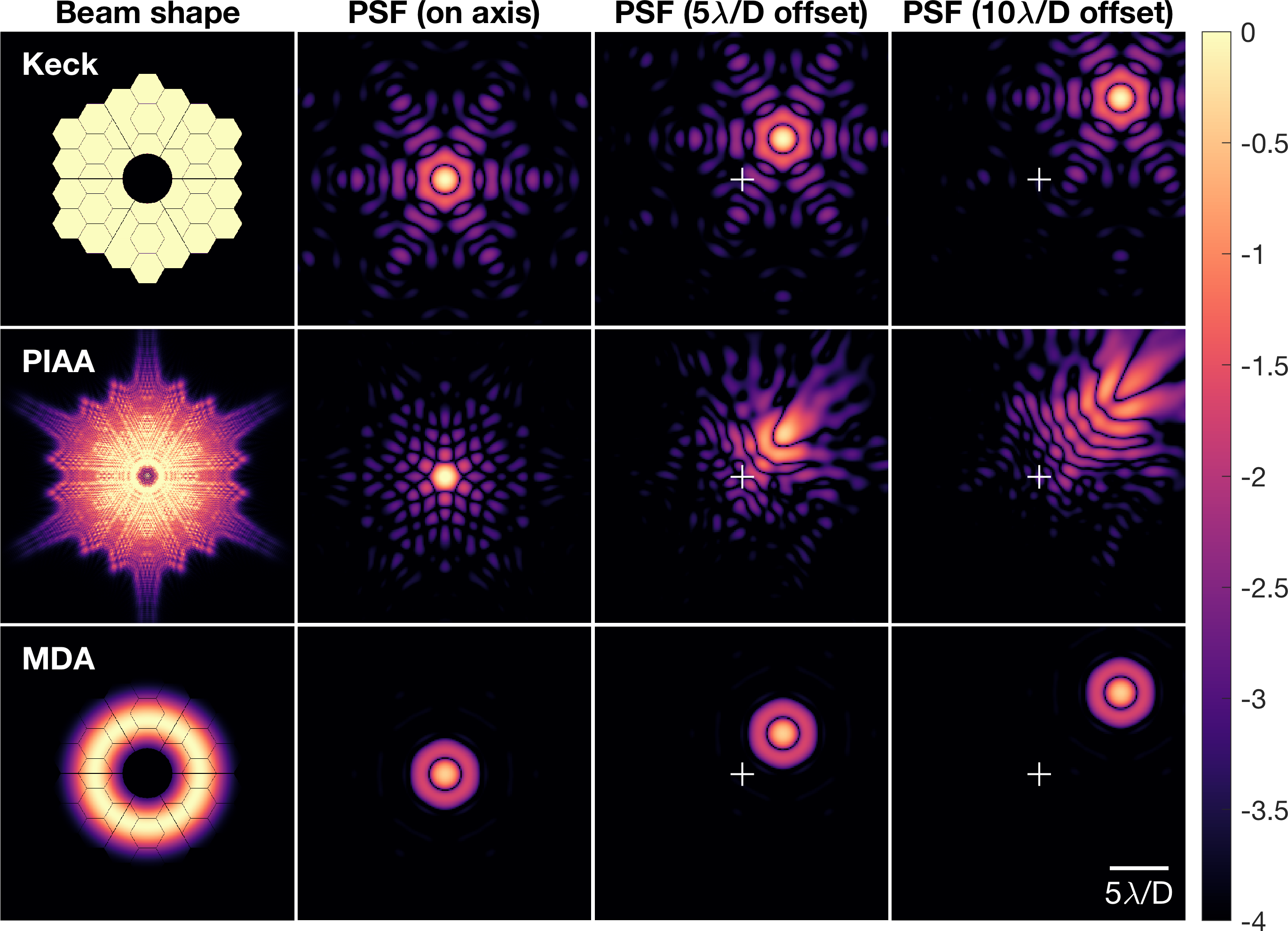}
\caption{Theoretical, monochromatic beam shapes and PSFs for the Keck telescope (Top row) without apodization, (Middle row) with the PIAA lenses, and (Bottom row) with the MDA. (Column 1) Beam intensity before focusing onto the SMF. (Column 2-4) Log base-10 PSF intensity for (Column 2) an on-axis source as well as sources with an angular separation of (Column 3) 5$\lambda/D$ and (Column 4) 10$\lambda/D$ from the optical axis (white cross). All are normalized to the peak intensity in the non-apodized case and using the corresponding central design wavelength (3.8~$\mu$m for the PIAA and 2.2$\mu$m for the MDA). \label{fig:all_psf}}
\end{figure*}

The KPIC PIAA optics consist of two aspheric lenses fabricated from CaF$_2$ because of its high transmission across the wavelength regime that KPIC operates ($K$ and $L$ bands; 2.0-4.2~$\mu$m). AR coatings for this spectral region were optimized and deposited on both sides of each optic to minimize Fresnel reflections. Figure~1(a) shows a ray diagram of the PIAA lenses, which are designed to re-distribute an annular intensity distribution, similar to the centrally-obscured Keck pupil, into a quasi-Gaussian profile. As a result, the PSF is a better match to the fundamental mode (LP$_{01}$) of a SMF and, therefore, the coupling efficiency is improved.

The lenses each consist of one flat surface (facing outwards) and one aspheric surface (facing inwards). As the light propagates from the first aspheric surface to the second 50~mm away, the collimated beam is remapped and recollimated. In this way the PIAA optic pair can be inserted into a collimated beam without modifying the downstream focusing optics. The sag profiles of the two lenses were designed by solving the differential equation in~\cite{guyon2003PIA}, which leverages Snell's law to achieve the desired remapping. Figure~\ref{fig:piaa}(b) and (c) show the fabricated lens pair while (d) and (e) show the cross-section of the rotationally-symmetric sag profiles. We chose the sag profiles to achieve the desired remapping function at the central wavelength in the $L^\prime$ band ($\lambda_0$~=~3.8~$\mu$m) because we expect the $S/N$ to be limited by thermal background in the longest wavelength filters. However, the PIAA can be used over the full KPIC range with slightly degraded coupling efficiency in the shorter wavelength filters.

One limitation of the PIAA optics is that the remapped intensity depends on the position of the source with respect to the optical axis. While the lenses are designed to provide a quasi-Gaussian beam shape for an on-axis source, the exiting wavefront deviates from this and becomes heavily aberrated when the source is off-axis by even a few resolution elements (i.e. $\lambda/D$, where $\lambda$ is the wavelength and $D$ is the beam diameter at the pupil). Figure~\ref{fig:all_psf} shows the beam shape and corresponding PSFs without any alteration (Fig.~\ref{fig:all_psf}, top row) and with (Fig.~\ref{fig:all_psf}, middle row) the PIAA optics. The PIAA modifies the PSF favorably in the case of an on-axis source by concentrating more light into the PSF core. However, the off-axis PSFs suffer from a strong coma aberration with a magnitude that increases with the the angular separation (Fig.~\ref{fig:all_psf} shows offsets of 5 and 10~$\lambda/D$). Due to the rotational symmetry of the PIAA design, the direction of the coma aberration will always point radially away from the core.

\begin{figure}[t]
    \centering
    \includegraphics[width = \linewidth]{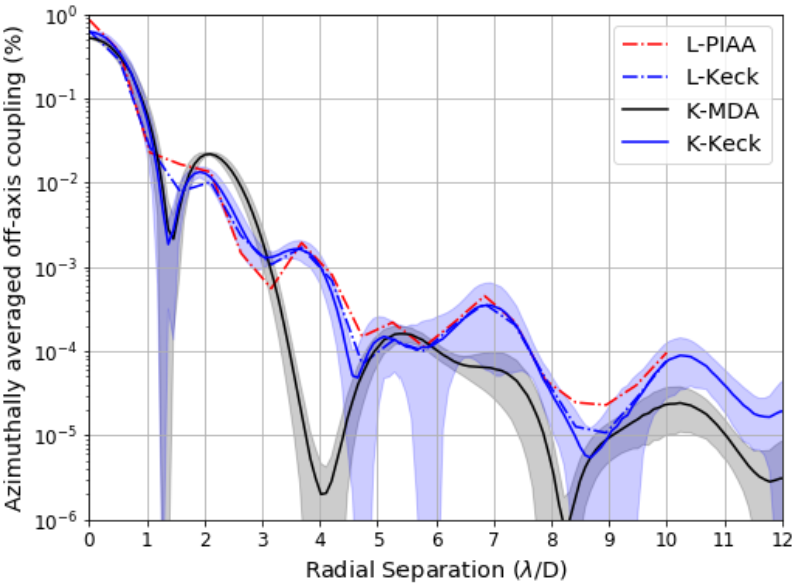}
    \caption{Azimuthal average of the coupling efficiency of an off-axis PSF shown in Fig.~\ref{fig:all_psf} at the central design wavelength and assuming no other aberrations. A coupling of 100\% means that all of the light of the PSF couples into the on-axis fiber. We can see that the PIAA does not significantly affect the off-axis coupling of the clear Keck pupil. We can also see that the MDA decreases the off-axis coupling from 3-12 $\lambda$/D, excluding two localized regions of 4.5-6 and 8.5-9.}
    \label{fig:offax_coup}
\end{figure}

To utilize PIAA optics for HDC, the planet must be aligned to the optical axis of the PIAA lenses to receive the coupling boost. The star would then be off axis, and for the separations depicted in the figure, would diffract a small amount of light onto the location of the planet marked by the white cross. This is where its important to note that Fig.~\ref{fig:all_psf} displays the intensity of the PSF, which does not necessarily indicate the amount of light that will couple into a SMF. So although there may be star light of non-zero intensity at the location of the planet fiber, we need to compute the coupling of that speckle pattern to the SMF to understand the stellar leakage term. Figure~\ref{fig:offax_coup} shows an azimuthally averaged line profile taken from a 2D coupling map computed for the PIAA, MDA and also with no optic in the beam (labelled Keck in the figure). It can be seen that the off-axis coupling response with the PIAA is very similar to that with a clear pupil, emphasizing that the PIAA does not increase stellar leakage. 
%, with approximately the same off-axis response as the non-apodized pupil.  

Where the PIAA makes use of transparent aspheric lenses, the MDA uses an array of opaque microdots to shape the pupil illumination to reduce the brightness of diffraction features around the PSF~\citep{Martinez2009a}. Since the MDA can be placed at, or near, the reimaged telescope pupil, the PSF is spatially-invariant; i.e. the PSF is approximately the same for the planet and star (see Fig.~\ref{fig:all_psf}, bottom row). The MDA for KPIC uses a semi-random pattern of 25~$\mu$m squares in a 200~nm thick chrome layer on an AR-coated CaF$_2$ substrate (seen in Fig.~\ref{fig:apod}). The size of the microdots was chosen to maintain a $\sim$10:1 ratio between the mask feature size and the wavelength of light to avoid anomalous diffraction effects \citep{Martinez2009b,zhang2018CMA}. This design is lossy because the chrome dots block light, creating a grayscale beam intensity distribution. The mask is optimized by maximizing the $S/N$ for a planet spectrum assuming the measurement is dominated by stellar photon noise. For computational convenience, rather than maximizing the $\eta_p/\sqrt{\eta_s}$ ratio, we minimize $\eta_s/\eta_p^2$, which is proportional to the required exposure time to achieve a given $S/N$ in a stellar photon noise limited regime.

To optimize the effective transmittance of the MDA, we assumed a polynomial radial apodization function with field amplitude
\begin{equation}
    A(r) = \sum_{j=0}^N c_j (r/a)^j,
    \label{eqn:apodtheory}
\end{equation}
where $c_j$ are constants, $r$ is the radial coordinate, and $a$ is the beam radius. In our simulations, we multiplied $A(r)$ by the expected Keck pupil function, which is not rotationally-symmetric. However, using a one-dimensional apodization function simplifies the optical system by not requiring a rotational alignment between the apodizer and the beam. For each set of $c_j$ coefficients, we computed the fraction of on-axis planet light that couples into the SMF, normalized to the total energy incident on the apodizer. This slightly modified definition of $\eta_p$ includes both losses in coupling efficiency and the semi-transparent design of grayscale mask. The effective $\eta_s$ was defined as the fraction of starlight that couples into the on-axis SMF when the star is imaged off-axis (see Fig.~\ref{fig:offax_coup}), averaged over planet-star separations of 3-15~$\lambda/D$. Using these metrics, we found the optimal apodization function by minimizing $\eta_s/\eta_p^2$ with a simplex algorithm (see Fig.~\ref{fig:all_psf}, bottom left). We repeated the optimization for various values of $N$ and opted to use $N=5$ since adding more terms did not significantly improve performance.

We converted the continuous apodization function into a binary microdot pattern using the Floyd-Steinberg error diffusion algorithm \citep{FloydSteinberg1976,Dorrer2007,zhang2018CMA} taking into account the relative transmission of the 200~nm thick chrome layer and the exposed substrate. Figure~\ref{fig:apod} shows the designed pattern and a photo of the fabricated MDA, which has approximately 500 microdots across the expected 12.3~mm beam size. The full coated region is 15~mm in diameter. Given that the $S/N$ is more likely to be stellar photon noise limited at the shorter wavelengths, we optimized the MDA for the central wavelength of $K$ band ($\lambda_0$~=~2.2~$\mu$m), but the MDA can technically be used with any KPIC filter. At the longest wavelengths, where thermal background tends to dominate, the MDA is not likely to improve performance because of the reduced throughput and potentially high emissivity of the opaque regions.

\begin{figure}[t]
    \centering
    \includegraphics[width = \linewidth]{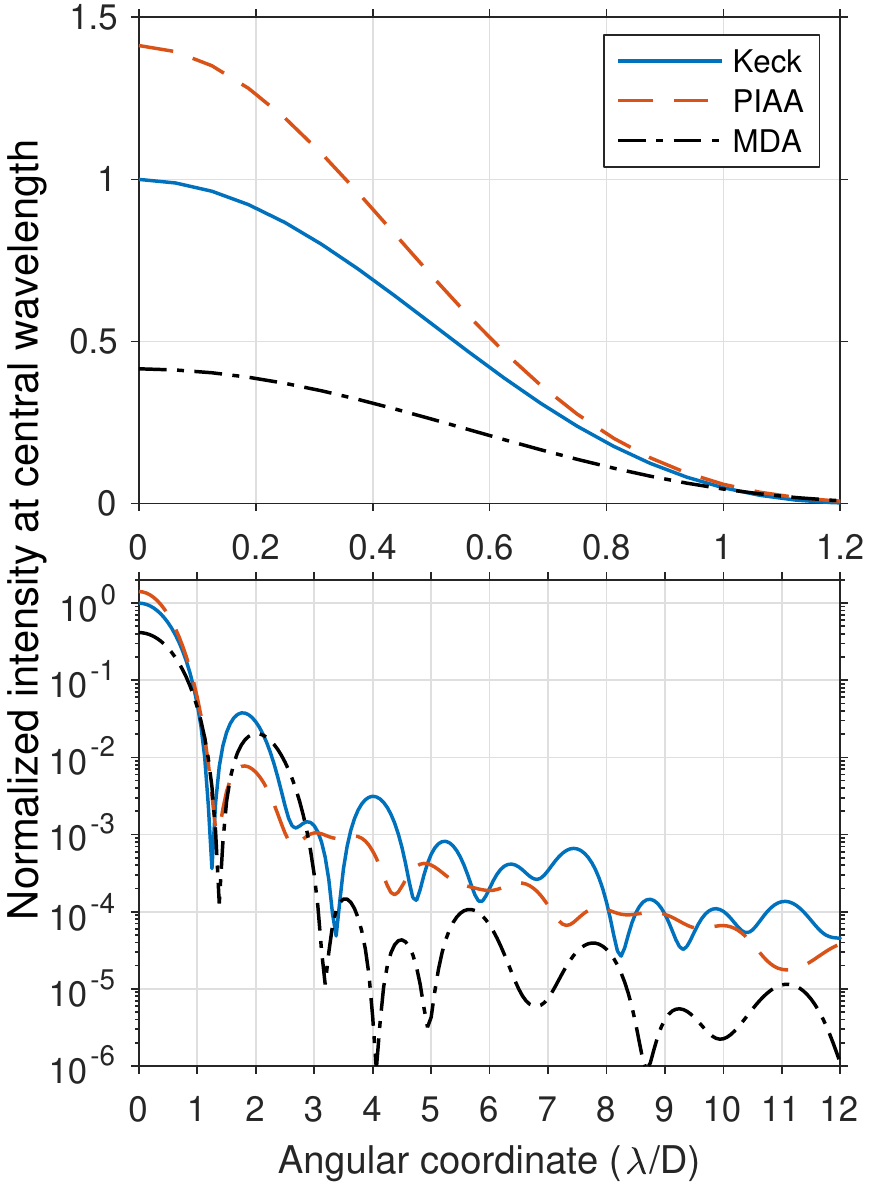}
    \caption{Azimuthal average of the on-axis PSFs shown in Fig.~\ref{fig:all_psf} at the central design wavelength and assuming no aberrations. The intensities in all cases are normalized to the peak of the Keck PSF. The top panel shows the intensity at the PSF core on a linear scale and the bottom panel shows the diffracted intensity further from the source on a log scale.}
    \label{fig:all_lineprof}
\end{figure}

Figure~\ref{fig:all_lineprof} shows azimuthally averaged line profiles of each on-axis PSF, normalized to the non-apodized case. The PIAA increases the flux in the core of the PSF, which leads to improved fiber coupling. On the other hand, the MDA reduces the diffraction from the star at angular separations beyond 3~$\lambda/D$, but at the cost of adding the semi-transparent optic and lowering the coupling efficiency. These technologies provide the ability to enhance the performance of KPIC in two different regimes. However, each has a significant downside. While the PIAA increases the throughput for an on-axis source, the source and PIAA must be carefully aligned to the optical axis of the system; the off-axis PSF becomes heavily aberrated. The MDA suppresses diffracted starlight and thereby reduces stellar photon noise in the measurement of the planet spectrum, but at the cost of overall throughput.

\begin{figure*}[t]
\centering
\includegraphics[width = \textwidth]{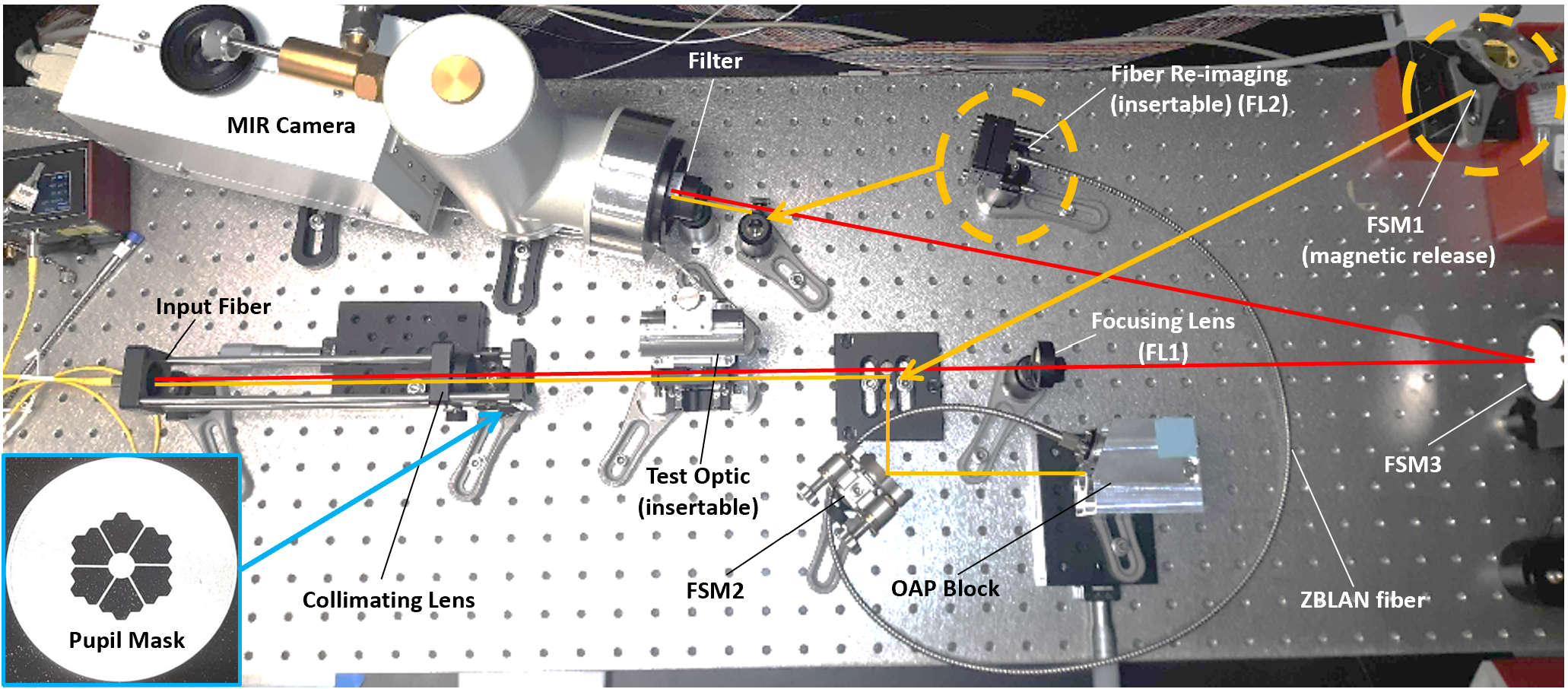}
\caption{Image of the experimental lab bench in the PSF imaging mode. The red line indicates the beam path for the PSF imaging mode. The orange line indicates the beam path for fiber coupling mode after the objects circled in orange have been inserted into the beam path. Inset in the bottom left corner is the pupil mask emulating the Keck telescope pupil. \label{fig:bench}}
\end{figure*}

\section{Experimental characterization} \label{sec:methods}
%We had the optics described in section 2 manufacturerd. Include the manufacturers?
%% Images of the PIAA.
%%% Talk about/image the mounts

%% Images of the apodizers
%%% Talk about/image the mounts
%%% Microscope images (25 um dots, defects are small in size, and etc)
%%% 

%___________________________________________________________________

%What parameters do we want to test? Spot shape, spot size, throughput, wavelength dependence on coupling etc. 
%% PSF
%%% Definition of apodization in terms of PSF
In this section we describe the experimental setup and tests carried out to characterize the fabricated devices. 

The key properties to examine for the two technologies are the PSF shape, the throughput, and the coupling efficiency. 
To specify, we use throughput in this paper to refer to the throughput of the individual optics and not the overall effect that the optic has on the system's throughput.
%In regards to the throughput, we will specifically be referring to the individual optics' throughputs and not the overall effect the optic has on the system throughput in this paper. 
%The individual optic's throughput is a result of Fresnel reflections, absorption, and reflections from the micro-dots in the case of the MDA. 
The throughput of the individual optics is less than 100\% as a result of Fresnel reflections, absorption, and reflections from the micro-dots in the case of the MDA.
Testing the throughput will help ensure that there were no errors in the manufacturing of the optics and that the AR coatings are performing according to spec. 
In addition, we will measure what effect these technologies have on the light coupled to a SMF on/off axis and compare this to simulations.
%In addition, we will measure what effect these technologies have on from what source and how much light gets into the SMF, and then we will compare this to what we expect from simulations.

%__________________________________________________________________
\subsection{Experimental Setup} \label{sec: setup}
%Describe experimental setup - figure to include a schematic and a lab photo - full annotated. 
Figure~\ref{fig:bench} shows the experimental setup with the PIAA mechanics in the beam.
The setup to characterize and measure the MDA is equivalent, but with the KPIC coronagraph module at the ``Test Optic" location.

% Describe the setup in words starting from beginning to end. (Thorlabs, BE1) was focused by a CaF2, f=1000 mm lens,
This testbed can be separated into 4 main components.
First, there is the ``telescope simulator," which collimates light injected by a SMF. 
The SMF can be swapped between a 2~$\mu$m laser and a blackbody light source that is single mode across the 2-5~$\mu$m wavelength region. The collimating optic is an AR coated CaF2 lens, with f=200 mm. Its distance from the light source can be adjusted with a translation stage to ensure that the light is collimated regardless of the wavelength. The beam then passes through a pupil mask, which imitates the Keck pupil. Given the long focal length of the collimating lens, the beam can be approximated as having a flat top illumination after passing through the mask. 

The second component to the testbed is the test optic. These are located in the beam immediately following the telescope simulator in collimated space.
Both the PIAA and MDA are mounted on stages that can repeatably move the optics in/out of the beam path. This allows for comparative measurements to be taken quickly between the apodized beam and the non-apodized beam.% with/without apodization.%% Optic can come in and out of the beam
% Can include images of the mounts here

Following the test optics is the third component, the coupling arm. 
This begins with a field-steering pickoff mirror (FSM1) that directs light toward the off-axis parabola (OAP) used to couple the light. %into the SMF.
FSM1 is mounted on a magnetic plate that can be removed and repeatably replaced to give quick access to the two main functionalities of the testbed. %%% OAP pickoff
This turning mirror is complemented by a second, fixed steering mirror (FSM2).  %%% OAP TT mirrors
The two in combination can translate the collimated beam and adjust its angle of incidence with respect to the OAP (inside of the OAP block). This allows for coupling to be efficiently optimized into the SMF. The OAP is a protected gold coated mirror with an off-axis focal length of 36.6 mm, which injects the light into a ZBLAN fiber (Le Verre Fluor\'e, MFD = $7.5~\mu$m at a $2.0~\mu$m wavelength, core diameter = $6.5~\mu$m, cladding diameter = $125~\mu$m). With the pupil mask diameter of 12.3~mm, this creates an $F\#=3.0$, which closely matches the measured ZBLAN fibers numerical aperture of 0.175 $\pm$ 0.01.

%% vs 1m focusing
Lastly, the setup also contains an imaging arm. The imaging arm has two modes, with the first taking the collimated output of the test optic and passing it through a f=1000~mm AR coated CaF2 focusing lens (FL1) to create an image. This arm is folded for convenience %%% 1m E coated lens
by a 2" silver mirror (FSM3) toward the camera. %%% Silver tip tilt mirror
The second imaging mode is re-imaging the output of the coupling fiber onto the camera. %%Coupling Re-imaging
The mounted fiber output is incident on a f=5.7 mm AR coated aspheric lens (FL2), which collects all of the light from the fiber and re-images it onto the camera. %%%5.7 mm D-coated aspheric lens
As seen in Fig.~\ref{fig:bench}, this reimaging assembly can be inserted and removed when switching from coupling mode to imaging mode.

An InSb camera (Merlin, Indigo) is used to carry out all imaging. It is sensitive from 1-5~$\mu$m, and needs to be liquid nitrogen cooled.
%Camera brand: Indigo. Label on top: "Ge window, SiO2 cold filter 1/10/2017
It has 240 x 318 pixels and a pixel pitch of 30~$\mu$m. %The pixel pitch that makes the simulations match the images
The PSF formed by the $F\#=81.3$ beam incident from FL1 is sampled by approximately 6 pixels per FWHM on this detector. 
Immediately in front of the camera, we place a spectral filter, used to both limit the background incident on the detector and to select the measurement band when using the blackbody light source. When we use the 2~$\mu$m laser and do K$_s$ measurements, a K$_s$-band filter (Asahi-Spectra) is used. For L band measurements, we use a 3-5 micron bandpass filter.
%a Thorlabs L filter is used. %%Filters (Nem can tell more)

%%%% Placement near the camera and the size of the PSF that would get generated means we don't need to worry too much about the exact focusing distance of this specific setup. It just needs to make a ~4-6px sized spot on the detector.

%__________________________________________________________________

\subsection{Procedures}
% Describe the experimental procedures. Break this into subsections - throughput, PSF analysis, and coupling efficiency - comment on camera tuning. 
%Everything is measured using the Indigo camera in order to give a consistent measurement basis.

Before the camera could be used to make quantitative measurements, it required us to linearize its response, maximize its dynamic range and do a one-point non-uniformity correction (NUC) flat fielding. The camera's exposure time and video offset were coarsely adjusted to match the expected flux levels for the experiments. Then, they were carefully tuned and the response of the camera to varying flux levels was compared to the reading of a power meter. This resulted in a linear camera response with the maximum possible dynamic range. The next step was to perform a one-point NUC to ensure a uniform response across the detector. 
This was done with the camera entrance window blocked by a black piece of metal, such that a uniform thermal background from the plate filled the camera's entire field-of-view, while running the NUC correction algorithm built into the camera.  %%%1pt Non-uniformity correction eliminates the need for flat-fielding as that's taken care of before data collection.
\subsubsection{PSF Imaging}
%%PSF
We recognize that the shape of the PSF is critical and that the faint wings can only be revealed at a high SNR. 
%When analyzing PSFs, the shape is critical and the faint wings can only be revealed at high SNR. 
With the beam of interest projected onto the camera, the exposure time and video offset were adjusted both to bring the background counts down to around 50-100 ADUs and to maximize the signal. The 1 point NUC was applied as outlined above. 
%The metal plate was removed. 
The light source was blocked and a cube of 100 frames that constitutes the background were collected. The light source was unblocked and a cube of 100 frames with signal was collected. Finally, the light source was blocked once again and another cube with 100 background frames was collected. The two sets of background frames, bracketing the signal frames, were used to construct a background that was as similar to the background when the signal was acquired as possible. %required because of how fast the detector properties can change as the LN2 evaporates.

%%% Integration time and video offset can be combined to maximize the signal/SNR of what is onset on the camera.
%%% Set T-Int and V/O for each image as the shape/deep structure is the most important thing in the PSF analysis.

\subsubsection{Throughput}
%%Throughput
To make quantitative measurements of the flux using the camera, aperture photometry/radiometry was conducted.
%%%%Lacking a quality powermeter/photodiode that is sensitive from 2 um through L-band, we realized that our MIR camera can serve as our power-reading source.
%Maximizing the SNR in the image is not as important in the throughput measurements. 
To obtain the throughput of a given optical element (e.g. PIAA or MDA), the flux from the PSF generated when the optic was in the beam was normalized by the flux of the PSF when the optic was out of the beam. 
As such, it was important to keep the exposure time and other camera settings constant between the two measurements to assure an accurate comparison. 

We take the same pattern of 100 background frames, 100 signal frames, and then another 100 background frames both when the optics were in and out of the beam. 

%%%%%We use the the Indigo camera to measure throughput 
%%%Do a PSF image with the optic in the beam and then with the optic out of the beam
%%%Keep the same Tint and V/O settings on the camera between the images.
%%%Conduct a radial flux measurement to account for failed background subtraction

\subsubsection{Coupling Efficiency}
%%Coupling Efficiency (the complicated one)
%%%Account for defocus in the PIAA and change the positions of the collimating and focusing lenses in a given band.
To determine the coupling efficiency to the ZBLAN optical fiber, we again compare the flux between two states: 1) light transmitted through the SMF re-imaged onto the detector and 2) the PSF that was directly incident on the fiber before being imaged onto the detector. These two states have non-common optics and so the throughputs of those optics had to be accounted for during the analysis. 

%The imaged PSF had a FWHM on the detector of $\sim4$~pixels.% This was large enough to ensure that insignificant amounts of flux are lost between pixels (prob cut this sentence, also it's not like we were gonna get a larger psf). 

To photometrically measure the coupling efficiency in a given band, we: 
\begin{enumerate}
\item Adjust the light source power, exposure time, and video offset to maximize the signal from the re-imaged coupling fiber,
\item Image, with backgrounds, the re-imaged fiber,
\item Carefully remove the re-imaging system and FSM1 to reconfigure into PSF imaging mode,
\item Image, with backgrounds, the PSF without adjusting any settings.
\end{enumerate}

After normalizing out the differential losses in each arm, we can take the ratio between the extracted fluxes and determine the coupling efficiency in isolation.

%_______________________________________________
% Data acquisition and analysis.
\subsection{Data Acquisition and Analysis}
%%%%%Can scrape the PIAA report for the basis of the radial flux measurement

There are multiple things we can do with the imaged PSF to analyze and quantify it's properties in comparison to the model. 
%In addition to calculating and plotting residuals, an example of a quantitative comparison would be 
For example, one can compute the overlap integral (Eqn ~\ref{eqn:couplingeff}) between the model and the experimental PSFs to assess their similarity. 
%This allows us to quantify the level of similarity between the simulation and the experimental result.
%
For all other forms of analysis (i.e. throughput or coupling), we must first extract the fluxes from the detector by carefully removing the background. % is measured through what we call a radial flux measurement. 
%When measuring the flux present on the detector, we need to track two sources of signal: the photons from the PSF that we want to measure, and from the thermal background. 
The PSF is localized to, at most, a region with a radius of 25 pixels. Meanwhile the thermal background is a non-zero bias across the frame, that can be assumed to be uniform following the 1 point NUC. When using the single-mode blackbody source, the signal is small, so this background subtraction must be done carefully.
%The reason this needs to be done so carefully is that this camera is sensitive out to 5~$\mu$m and its gains seems to change on a very fast timescale (possibly due to LN2 boil off). %  however, the change in thermal background of the detector occurs at such a high rate at long exposure times that there will still be some constant background even with background subtraction. 
To accurately only extract the flux from the PSF, we follow a two step process: 1) subtract a background frame collected close in time to the data and 2) model out the remainder of the background.

To achieve step 2, we take measurements of the total enclosed aperture flux at increasing aperture radii. Beyond the extent of the PSF, the encircled flux increases or decreases linearly (for a uniform background) with the total enclosed pixels, $F_{internal} = F_{PSF} + \alpha \cdot n_{pix}$, where $\alpha$ is the average background flux. 
Thus, in order to get $F_{PSF}$, we fit the linear part of the curve and solve to determine the y-intercept, which is the background free flux. 
The uncertainty in this measurement is the relevant value from the covariance matrix of the fit.

%%PSF centered, normalized, log to compare to model PSF

%%Throughput = beam_in_flux / beam_out_flux
With the flux's extracted, the throughput can next be calculated. It is the ratio of the fluxes: 
$\xi = F_{in}/F_{out}$
%%%Include radial flux measurement here

The coupling efficiency is mathematically very similar to the throughput, the ratio of the flux coupled into the mode of the SMF and the flux incident on the fiber face. However, there is more involved with getting to this step. As was stated in the procedure, there are many non-common optics between the fiber re-imaging block and the focusing arm of the bench. 
Once those are accounted for, the coupling efficiency is as well: 
$\eta = F_{coup}/F_{in}$
%%Coupling calculated experimentally with radial flux measurements
%%%Coupling efficiency = flux propagated through fiber / flux incident on the fiber
%%%% Can calculate flux propagated through fiber with RFM corrected for the fresnel reflections of the fiber surface and the throughput of the 5.7 mm lens.
%%%% Can calculate flux incident on the fiber with a RFM of the PSF, corrected for the throughputs of the 2inch silver mirror, the gold FSM pickoffs, and the gold OAP.

The key to qualifying the experimental measurements is to be able to compare them to accurate models. To ensure we get the highest degree of agreement, we replicate the parameters of Fig.~\ref{fig:bench} in a simulator. There are uncertainties in some of the physical parameters in the coupling arm of the setup: the NA of the coupling fiber and the effective focal length of the coupling OAP are known only to $\sim5\%$. These uncertainties were accounted for in simulations and resulted in a range of expected values.
%This will manifest as a range of expected values, rather than a definitive expected value of coupling efficiency.

\section{Results} \label{sec:results}
Here we summarize the measurements of the parameters outlined in the methods section and compare them to simulations.

\subsection{PSF profiles}
\begin{figure}[h]
\centering
\includegraphics[width = 0.45\textwidth]{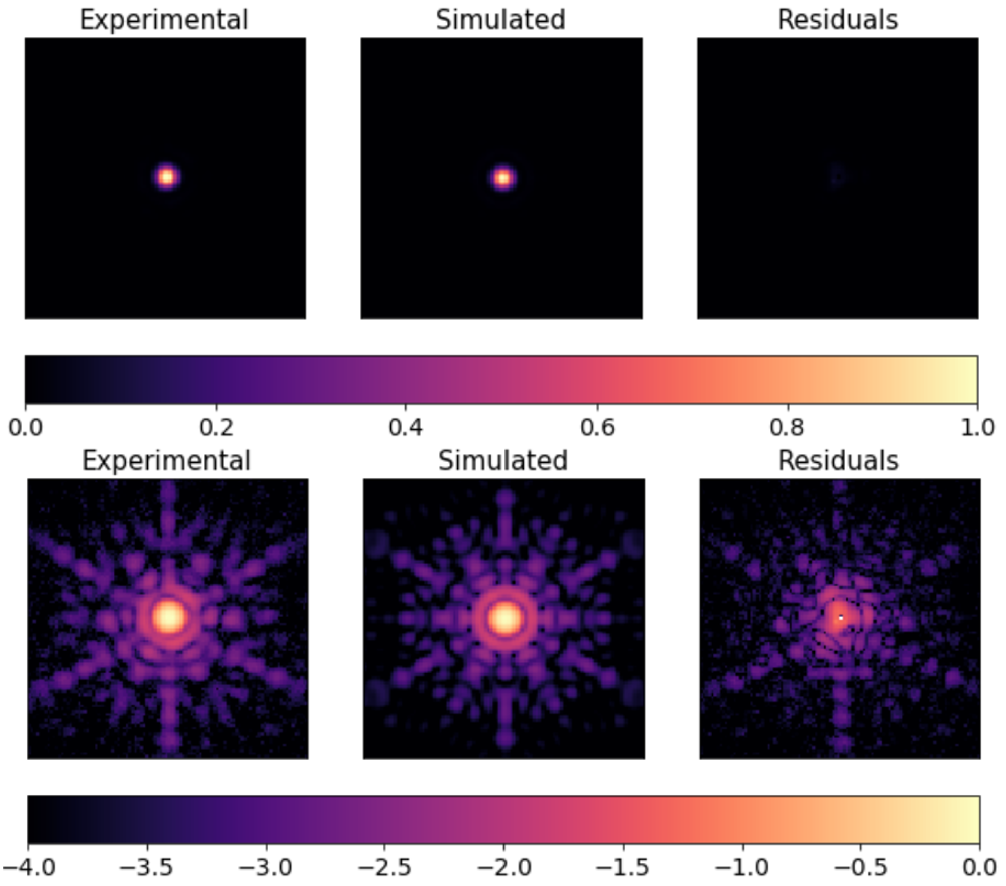}
\caption{PSFs for the PIAA shown in both linear (top) and a base-10 logarithmic (bottom) space with the same spatial extent. (Left) Experimental, (Center) Simulated, (Right) Residual of difference between the model subtracted from the experimental PSF. The linear scaling residuals emphasize the high degree of agreement between the model and experiment, while the logarithmic scaling shows the structural differences between the two. All PSFs are shown at $2~\mu$m. \label{fig:piaa_psf}}
\end{figure}

\begin{figure}[h]
\centering
\includegraphics[width = 0.45\textwidth]{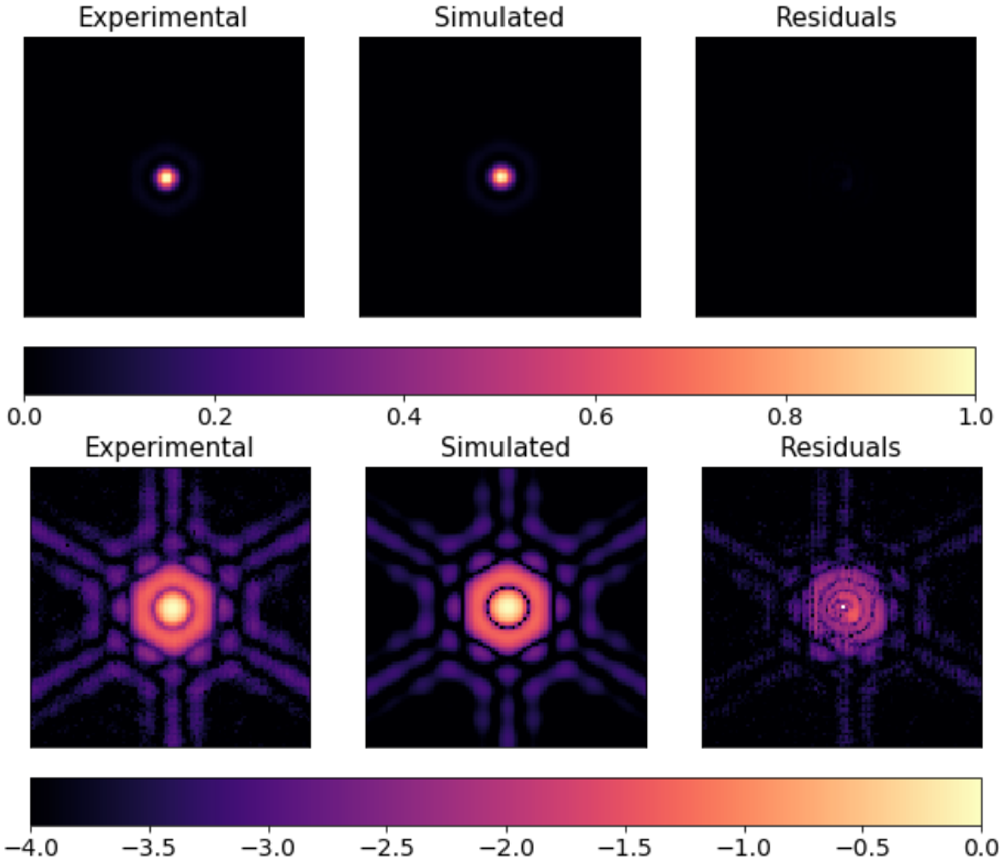}
\caption{PSFs for the MDA shown in both linear (top) and a base-10 logarithmic (bottom) space with the same spatial extent. (Left) Experimental, (Center) Simulated, (Right) Residual of difference between the model subtracted from the experimental PSF. The linear scaling residuals emphasize the high degree of agreement between the model and experiment, while the logarithmic scaling shows the structural differences between the two. All PSFs are shown at $2~\mu$m.  \label{fig:apod_psf}}
\end{figure}

The critical feature for both technologies is the structure of the PSFs and specifically the faint structure in the wings. Only the 2~$\mu$m laser had enough power to reveal the faint structure on the MIR camera. Figure~\ref{fig:piaa_psf} and \ref{fig:apod_psf}, present the PSFs obtained from the Indigo camera side-by-side with the corresponding simulations and the residual difference between the two. This enables qualitative comparison between the two, which is important to building confidence in both the manufacturing and design process.

\begin{figure}[h]
\centering
\includegraphics[width = 0.45\textwidth]{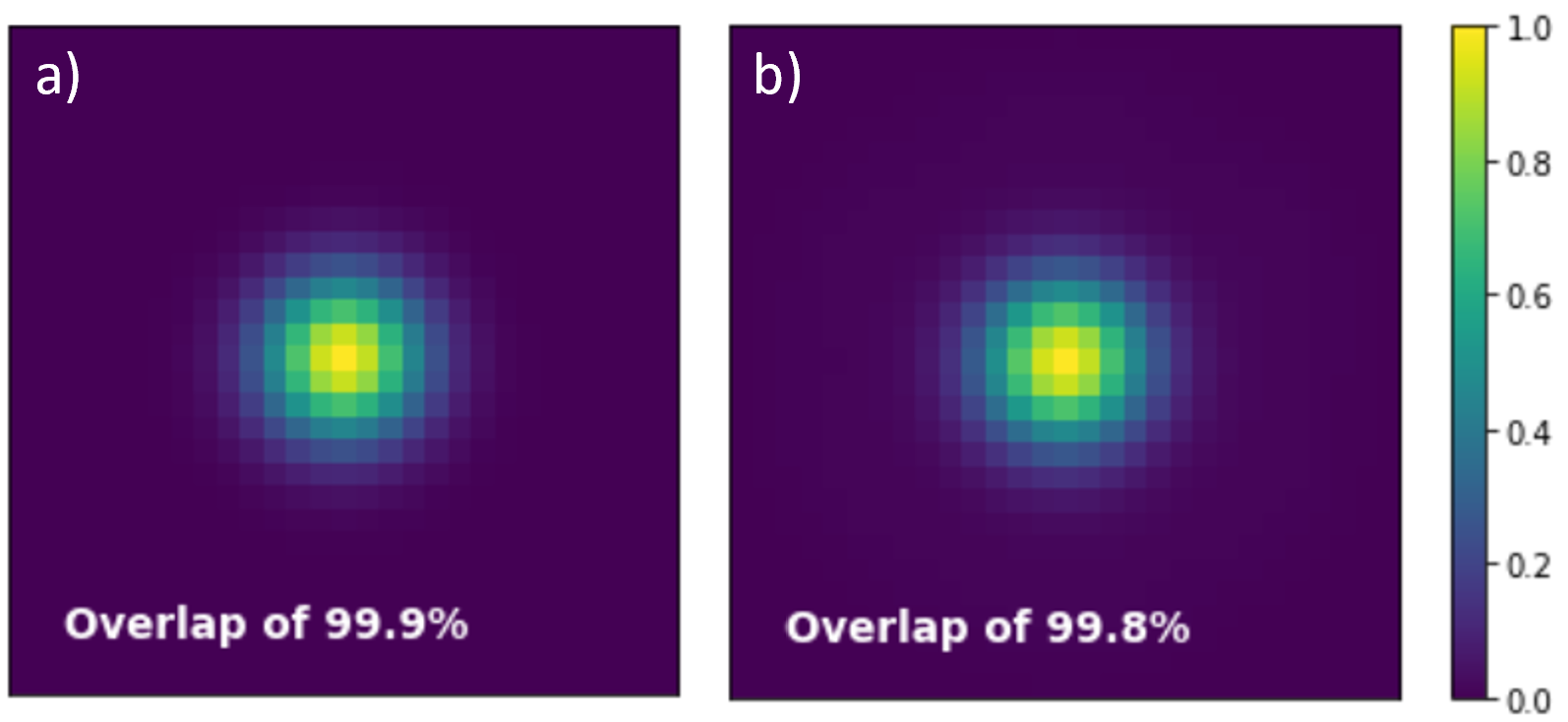}
\caption{Overlap integral maps between experimental and modelled PSFs for (Left) PIAA and (Right) MDA. The figures show a high level of consistency between the imaged PSFs and the simulated ones. \label{fig:overlap}}
\end{figure}

It can be seen from both figures, that there is a strong similarity between the experimental and simulated PSFs. The right hand panel in both figures shows that the residuals of the difference between the model and the experimental PSFs are very low, with a peak around the $10^{-2}$ or 1\% intensity level. Further, we can also use the overlap integral as a quantitative comparison tool for the PSFs. To account for misalignment's in the peaks of the two PSFs, we scan the model PSF across the experimental PSF in 2D and compute an overlap integral map as shown in Fig.~\ref{fig:overlap}. From the overlap integral maps, we can see that both technologies have a peak greater than 99.8\%, indicating a high level of consistency between the experimental and simulated PSFs, validating the design and manufacturing process.

%A benefit of running the convolution is that now we can see how much lateral displacement from the on-axis point can be tolerated before it stops acting as an on-axis PSF. In the discussion section, we'll take some time to show how the on-axis vs the off-axis PSF is an important consideration for when to use which technology.

\begin{figure}[h]
\centering
\includegraphics[width = 0.45\textwidth]{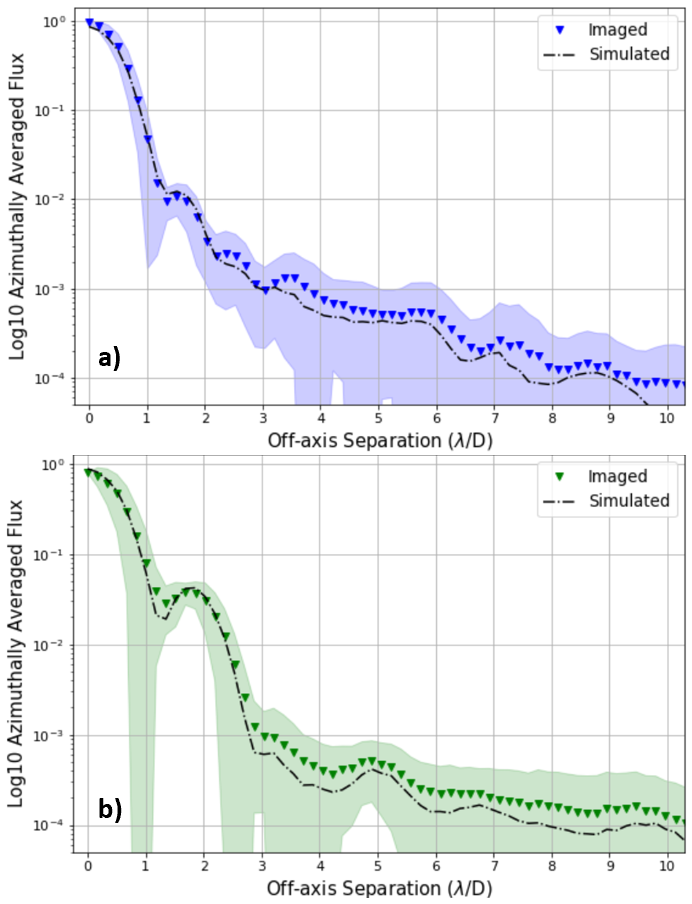}
\caption{Azimuthally averaged line profile of the a) PIAA PSF and b) MDA PSF. The shaded region accounts for the uncertainty in the measurements. \label{fig:exp_lineprof}}
\end{figure}

We can also compare the azimuthally averaged line profiles of the PSFs. By over plotting the line profile of the experimental PSF with the simulated PSF, we can identify locations where the flux is unexpectedly high or low. In Fig.~\ref{fig:exp_lineprof}, we show these line profiles of the two technologies. The shaded regions indicate the standard deviation in the azimuthal direction of the imaged PSF. We can see that both the PIAA and MDA closely follow the simulated curves, once again emphasizing the similarity between the design and the characterized devices.

\subsection{Throughput}

\begin{figure}[h]
\centering
\includegraphics[width = 0.45\textwidth]{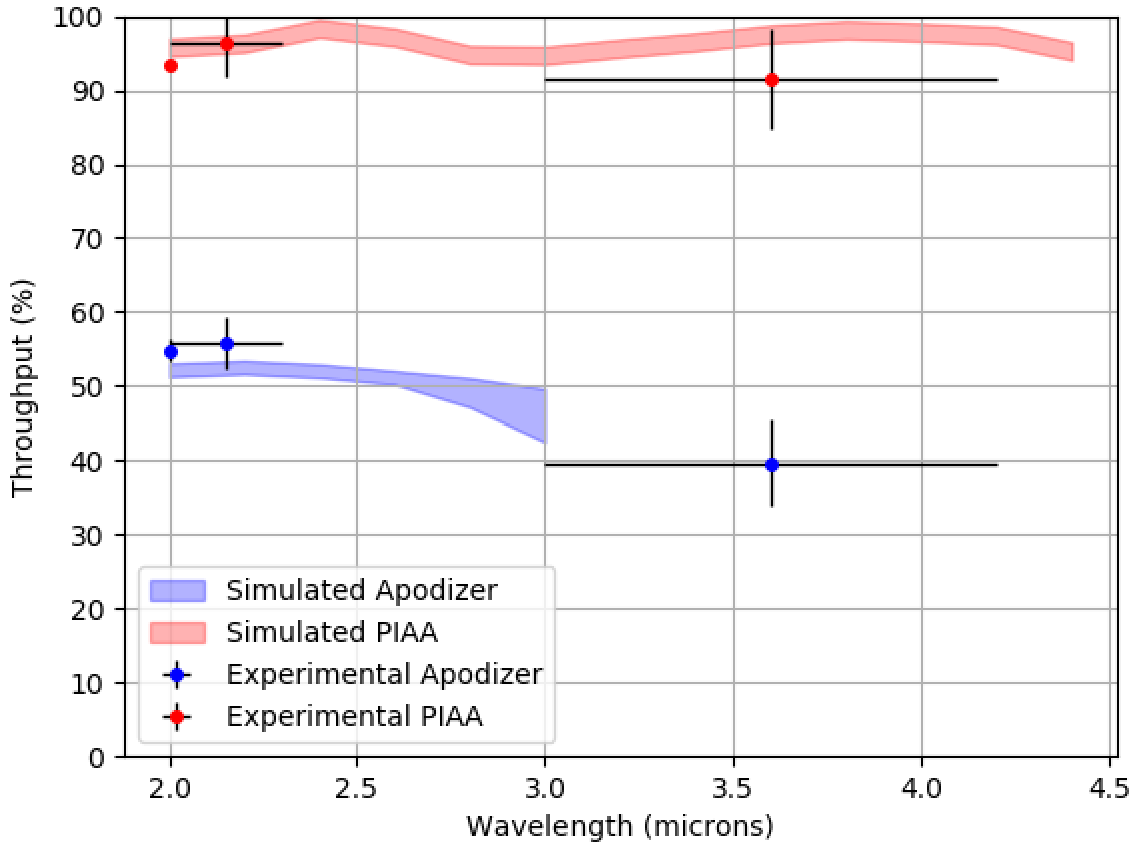}
\caption{A comparison of throughput curves between simulations and our experimental measurements for the PIAA and MDA optics. \label{fig:all_thru}}
\end{figure}

The throughput is defined as $\xi = F_{out}/F_{in}$, where $F_{in}$ is the flux at the input to the test optic and $F_{out}$ is the flux at the output of the test optic. The simulations were carried out in discrete monochromatic bands, while the experimental measurements, aside from the 2~$\mu$m laser, were taken using a broadband black-body light source combined with bandpass filters. The simulated and measured throughputs are shown in Fig.~\ref{fig:all_thru}, where the horizontal error bars represent the bandwidth of the lab filters.

As can be seen from the figure, both technologies met the expected throughput to within uncertainty. The PIAA is designed to reshape the pupil in a lossless fashion as described in Section~\ref{sec:tech}. The throughput is well above 90\% in both K and L bands as expected, 
and any systematic losses can be easily explained through residual reflections across the four AR coated surfaces.
%and the losses are fully explained by residual reflections from imperfect AR coatings (4 surfaces in total).

We can see that the MDA had a throughput consistent with simulations of $\sim50\%$ in K-band. However, we see it begin to roll off in the experimental results in the L band. This is due to the AR coating on the MDA not being optimized for L band, because the losses associated with the MDA reduce the L-band flux to the point where the thermal background dominates. Therefore, the MDA was not expected to be scientifically useful in L band and hence the coating was not optimized in this range.%an issue. 
%In Section~\ref{sec:discuss_apod} we elaborate on how the MDA was not optimized in these wavelengths.
%This is because it was not intended to operate in this range as will be justified in Section~\ref{sec:discuss_apod}. 

\subsection{Coupling Efficiency}
\begin{figure}[h]
\centering
\includegraphics[width = 0.45\textwidth]{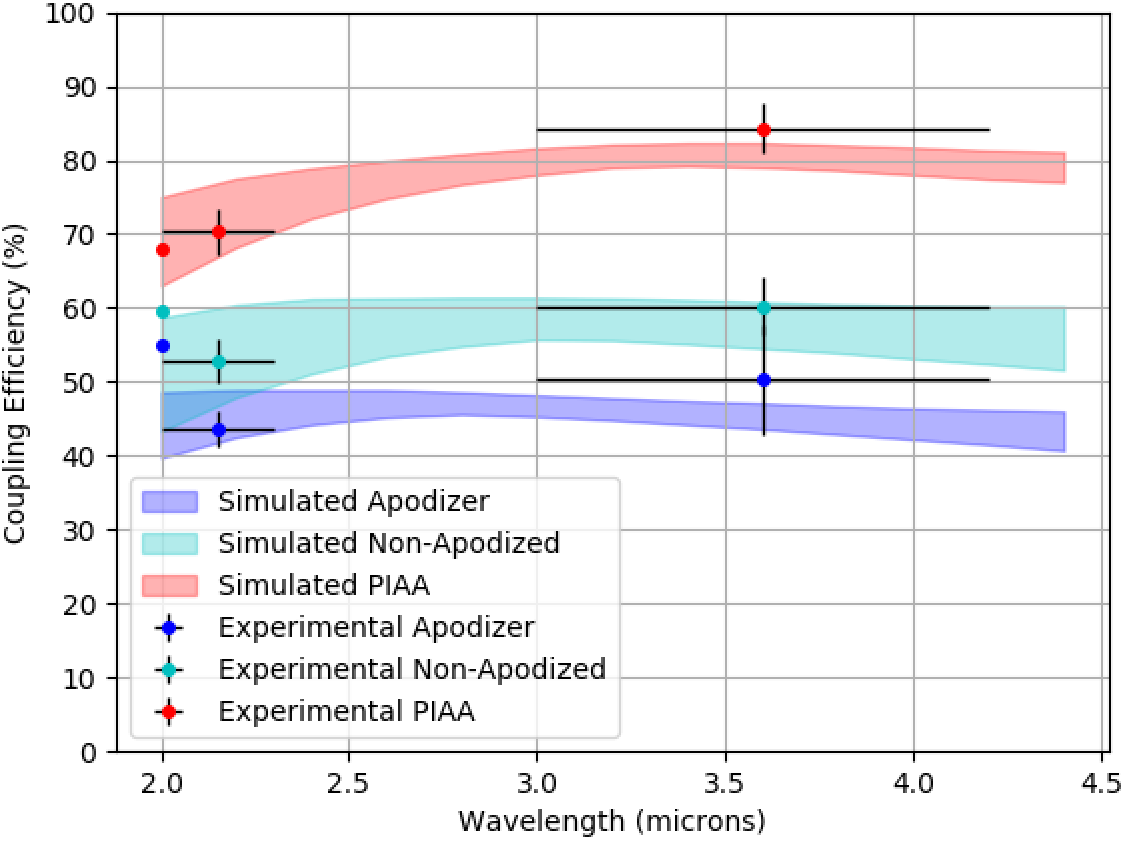}
\caption{A comparison of coupling efficiency curves between the simulation and our experimental measurements for both test optics. \label{fig:all_coup}}
\end{figure}

The simulations for the coupling efficiency were also produced in discrete monochromatic bands, and compared to the polychromatic results obtained with a filtered black-body light source in the laboratory. The coupling efficiencies for the two technologies are shown in Fig.~\ref{fig:all_coup}. 
The expected coupling for the non-apodized pupil is overlaid on these figures for reference. 

It can be seen that the coupling for the PIAA optics matches our simulation well. We can also see that the coupling for the MDA matches the simulation quite well, except for the measurement with the 2 $\mu$m laser. We are unsure why this data point is off, but believe it has to do with the increased coherence of the 2 $\mu$m laser and interference effects generated by the MDA. To ensure that this was not merely an effect of the camera, the measurement was repeated by placing a Thorlabs S148c power meter near the final focal plane and making the same flux comparison as the camera. The power meter confirmed the higher than expected coupling when using the MDA and the 2 $\mu$m laser beam. While this is not well understood, it is not an undesirable result in the context of KPIC. 

%ASIDE: I believe this is a result of some weird resonance/phase stuff that is happening between the coherent laser and the microdots. Rebecca and Gary saw some weird looking stuff happening with the phase in the microdot paper, so there might be some equivalent phase issue that's happening in this optic that is artificially increasing the coupling -BC\\

From the figure it is clear that the PIAA boosts the coupling from $\sim60\%$ in K and L bands without any additional optics to $\sim70\%$ in K and $84\%$ in L band. In comparison, the MDA reduces the coupling in K band to $\sim48\%$ and in L band to $45\%$. It should be made clear that these results are for the Keck like pupil mask in our experiments, which has oversized spiders, and at the Keck telescope, the coupling will be $3-4\%$ greater in all cases. 

Looking at the impact of the optics at the system level by combining the throughput and coupling efficiency, the PIAA has a total system throughput of $66\%$ in K band and $77\%$ in L band while the MDA is closer to $\sim24\%$ in both K and L bands. The impact of these efficiencies will be discussed in the following section.

\section{Discussion} \label{sec:discussion}
%We should discuss how and when to use each of the technologies
It is clear from the results that the PIAA and MDA have different characteristics that will be useful in different contexts, so we will address them individually.

\subsection{Beam Shaping Optics} \label{sec:discuss_piaa}
The PIAA are lossless beam-shaping optics. 
We can see from the Ks and L band measurements that there is an increase in coupling efficiency, from $\sim60\%$ without the PIAA to $\sim75\%$ with PIAA in K band and $\sim85\%$ in L band, where the lenses were optimized to operate. This boost in coupling is only present for the on-axis source as PIAA optics suffer strong aberrations off-axis as shown in Section~\ref{sec:tech} above. The lenses have a small loss ($\sim8\%$), which is equal for both the star and the planet. For KPIC to utilize these optics effectively, the known planet will need to be aligned with the optical axis of the PIAA optics, i.e, on-axis. This will result in a throughput increase for the planet light (improved coupling), and a reduction in the starlight (losses from optics and aberrations for off-axis star). Although the PIAA optics may scatter some of the star light onto the location of the planet fiber, for small angular separations between the star/planet as shown in Fig.~\ref{fig:all_psf}, the coupling to the fiber will be extremely low. In this way, the PIAA optics can reduce the exposure time to reach a given SNR, which will also allow fainter objects to be observed.

\subsection{Micro-dot Apodizer} \label{sec:discuss_apod}
The MDA's effect is more subtle. The role of the MDA is to decrease the amount of contaminating light from an off-axis source that gets coupled into the SMF. As can be seen from the line profiles (Fig.~\ref{fig:all_lineprof}), the PSF from the apodizer has nearly an order of magnitude less flux between 3 and 12 $\lambda$/D compared to the PSF from a non-apodized pupil. So, in situations where the light from the on-axis host star is drowning out the signal from the off-axis planet, using the MDA can help isolate the planet light. One drawback of the MDA is, to achieve such starlight suppression, the apodizer must decrease the throughput of the on-axis (star) and off-axis (planet) objects to $\sim50\%$. Furthermore, the coupling is $48\%$, lower than the native PSF and significantly lower than when using the PIAA optics. Therefore, one must carefully consider when to use this technology. 
%Apodizer has lower planet coupling, but better starlight suppression. So may be better inwards of 5 lambda/D
%PIAA is lossless, and offers improved planet throughput, so can be used in L band, where thermal background is high

\subsection{Exposure Time Calculation} \label{sec:ETC}
\begin{figure*}[ht]
\centering
\includegraphics[width = 0.99\textwidth]{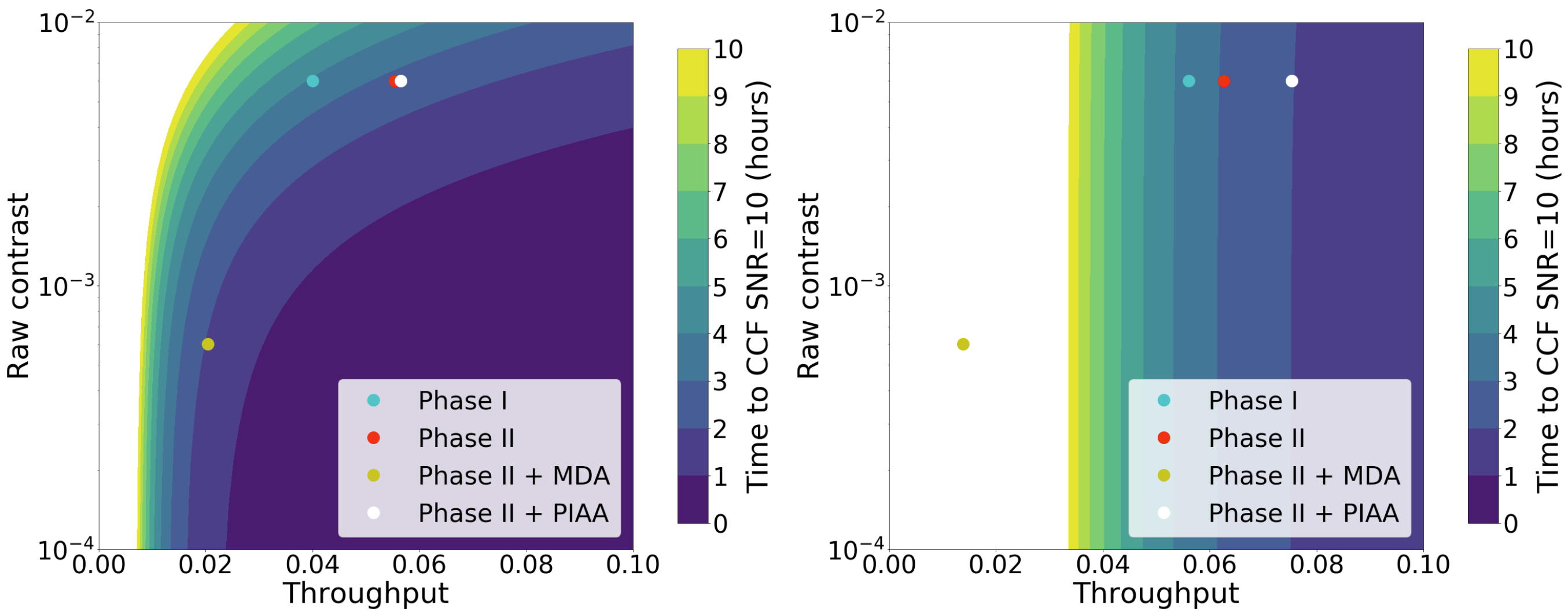}
\caption{Exposure time contours for S/N=10 detection of 51 Eri b using the different technologies in K-band and L-band in units of hours. Left, we can see in the K-band how the MDA decreases the planet's throughput, but the resultant increase in star contrast creates a net improvement in detection exposure time. Right, we can see how the increased throughput of the PIAA helps to decrease the detection exposure time. \label{fig:51_Eri}}
\end{figure*}
 
In this section we look at the impact of the two technologies on the exposure time of a simulated observing scenario of the planetary system 51 Eridani b, with KPIC, in both K and L band. Although the star is typically on-axis during acquisition, a tip/tilt mirror in the injection unit is adjusted in the final steps of acquisition to put the known planet on-axis and align it with the SMF. In this scenario, the exposure time ($\tau_{exp}$) of an observation to reach a given SNR is proportional to the coupling efficiency of the stars PSF ($\eta_{s}$) and approximately inversely proportional to the square of the coupling efficiency of the planet PSF ($\eta_{p}$): 
$\tau_{exp} \propto \eta_{s} / \eta_{p}^2$

To see how each technology effects the exposure time, the properties of KPIC were included in a realistic exposure time calculator (ETC). $\eta_{s}$ and $\eta_{p}$ were scanned across a grid of reasonable values for the case of observing 51 Eridani b. Figure~\ref{fig:51_Eri} shows the contour maps of exposure time vs raw contrast and throughput to achieve a cross-correlation function (CCF) S/N=10. On these figures, discrete points can be seen for the case of the native PSF (red dots), the PIAA (white dots) and the MDA (gold dots) PSFs. In addition, the cyan dots highlight the phase I level of performance of KPIC on-sky before deploying these technologies in phase II.  

It can be seen that in the case of 51 Eridani b, the PIAA optics offer little improvement in K-band compared to the expected phase II performance, while the MDA can reduce the exposure time by $\sim33\%$ from 3 to 2 hrs. Despite the fact the MDA decreases the planet's throughput, it also suppresses the star light, improving contrast and the net effect is an improvement in the detection exposure time. However, in L-band, which is dominated by thermal background, the PIAA optics offer the advantages in exposure time, reducing the exposure time of the expected phase II system by $\sim33\%$ from 3 to 2 hrs, while the MDA should be avoided. Factors of $33\%$ are significant when it comes to observing time on 8-10 m class telescopes.

This demonstrates the relative strengths and weaknesses of the two technologies, and their applicability to a given observing scenarios depends on the waveband of observation, the flux ratio between the star/planet and their separation.

\section{Conclusion} \label{sec:conclusions}
We have demonstrated the design, fabrication, characterization and accurate simulation of both PIAA and MDA optics in the context of HDC. The optics were optimized to operate in the K and L bands, which overlaps with the operating spectral bands of the KPIC instrument. The PSFs, throughput and coupling efficiencies all matched expectations from simulations, providing a high level of confidence in both the simulation tools and manufacturing capabilities. In a simulated observing scenario of 51 Eridani b, it was determined that the MDA would offer a reduction in exposure time of $\sim33\%$ in K-band, while the PIAA would offer a similar level of reduction for L-band observations. Such reductions in exposure time are important on 8-10 m class telescopes and are necessary to be able to target even fainter objects in future. The application of a particular technology to a given observation will depend on waveband of choice, the flux ratio of the star/planet and their angular separation, and needs to be determined on a case-by-case basis.

\acknowledgments
This work was supported by the Heising-Simons Foundation through grants \#2019-1312, \#2017-318 and \#2015-129. Support was provided by the Simons Foundation grant "Planetary Context of Habitability and Exobiology”

G. Ruane was supported by an NSF Astronomy and Astrophysics Postdoctoral Fellowship under award AST-1602444. 
W. M. Keck Observatory is operated as a scientific partnership among the California Institute of Technology, the University of California, and the National Aeronautics and Space Administration (NASA). The Observatory was made possible by the generous financial support of the W. M. Keck Foundation. The authors wish to recognize and acknowledge the very significant cultural role and reverence that the summit of Maunakea has always had within the indigenous Hawaiian community. We are most fortunate to have the opportunity to conduct observations from this mountain. Part of this work was carried out at the Jet Propulsion Laboratory, California Institute of Technology, under contract with NASA.

%\vspace{3mm}
\facilities{KeckII}

\vspace{1mm}

%% For this sample we use BibTeX plus aasjournals.bst to generate the
%% the bibliography. The sample63.bib file was populated from ADS. To
%% get the citations to show in the compiled file do the following:
%%
%% pdflatex sample63.tex
%% bibtext sample63
%% pdflatex sample63.tex
%% pdflatex sample63.tex

\bibliography{main_kpic}{}

\begin{thebibliography}{}
\expandafter\ifx\csname natexlab\endcsname\relax\def\natexlab#1{#1}\fi
\providecommand{\url}[1]{\href{#1}{#1}}
\providecommand{\dodoi}[1]{doi:~\href{http://doi.org/#1}{\nolinkurl{#1}}}
\providecommand{\doeprint}[1]{\href{http://ascl.net/#1}{\nolinkurl{http://ascl.net/#1}}}
\providecommand{\doarXiv}[1]{\href{https://arxiv.org/abs/#1}{\nolinkurl{https://arxiv.org/abs/#1}}}

\bibitem[{{Barman} {et~al.}(2015){Barman}, {Konopacky}, {Macintosh}, \&
  {Marois}}]{barman2015SDW}
{Barman}, T.~S., {Konopacky}, Q.~M., {Macintosh}, B., \& {Marois}, C. 2015,
  \apj, 804, 61, \dodoi{10.1088/0004-637X/804/1/61}

\bibitem[{{Beuzit} {et~al.}(2019){Beuzit}, {Vigan}, {Mouillet}, {Dohlen},
  {Gratton}, {Boccaletti}, {Sauvage}, {Schmid}, {Langlois}, {Petit},
  {Baruffolo}, {Feldt}, {Milli}, {Wahhaj}, {Abe}, {Anselmi}, {Antichi},
  {Barette}, {Baudrand}, {Baudoz}, {Bazzon}, {Bernardi}, {Blanchard}, {Brast},
  {Bruno}, {Buey}, {Carbillet}, {Carle}, {Cascone}, {Chapron}, {Charton},
  {Chauvin}, {Claudi}, {Costille}, {De Caprio}, {de Boer}, {Delboulb{\'e}},
  {Desidera}, {Dominik}, {Downing}, {Dupuis}, {Fabron}, {Fantinel}, {Farisato},
  {Feautrier}, {Fedrigo}, {Fusco}, {Gigan}, {Ginski}, {Girard}, {Giro},
  {Gisler}, {Gluck}, {Gry}, {Henning}, {Hubin}, {Hugot}, {Incorvaia}, {Jaquet},
  {Kasper}, {Lagadec}, {Lagrange}, {Le Coroller}, {Le Mignant}, {Le Ruyet},
  {Lessio}, {Lizon}, {Llored}, {Lundin}, {Madec}, {Magnard}, {Marteaud},
  {Martinez}, {Maurel}, {M{\'e}nard}, {Mesa}, {M{\"o}ller-Nilsson}, {Moulin},
  {Moutou}, {Orign{\'e}}, {Parisot}, {Pavlov}, {Perret}, {Pragt}, {Puget},
  {Rabou}, {Ramos}, {Reess}, {Rigal}, {Rochat}, {Roelfsema}, {Rousset}, {Roux},
  {Saisse}, {Salasnich}, {Santambrogio}, {Scuderi}, {Segransan}, {Sevin},
  {Siebenmorgen}, {Soenke}, {Stadler}, {Suarez}, {Tiph{\`e}ne}, {Turatto},
  {Udry}, {Vakili}, {Waters}, {Weber}, {Wildi}, {Zins}, \&
  {Zurlo}}]{beuzit2019SPH}
{Beuzit}, J.~L., {Vigan}, A., {Mouillet}, D., {et~al.} 2019, \aap, 631, A155,
  \dodoi{10.1051/0004-6361/201935251}

\bibitem[{{Bottom} {et~al.}(2016){Bottom}, {Femenia}, {Huby}, {Mawet},
  {Dekany}, {Milburn}, \& {Serabyn}}]{bottom2016SNW}
{Bottom}, M., {Femenia}, B., {Huby}, E., {et~al.} 2016, in \procspie, Vol.
  9909, Adaptive Optics Systems V, 990955

\bibitem[{{Crass} {et~al.}(2020){Crass}, {Bechter}, {Sands}, {King},
  {Ketterer}, {Engstrom}, {Hamper}, {Kopon}, {Smous}, {Crepp}, {Montoya},
  {Durney}, {Cavalieri}, {Reynolds}, {Vansickle}, {Onuma}, {Thomes}, {Mullin},
  {Shelton}, {Wallace}, {Bechter}, {Vaz}, {Power}, {Rahmer}, \&
  {Ertel}}]{crass2020-FDO}
{Crass}, J., {Bechter}, A., {Sands}, B., {et~al.} 2020, \mnras,
  \dodoi{10.1093/mnras/staa3355}

\bibitem[{{Currie} {et~al.}(2017){Currie}, {Guyon}, {Tamura}, {Kudo},
  {Jovanovic}, {Lozi}, {Schlieder}, {Brandt}, {Kuhn}, {Serabyn}, {Janson},
  {Carson}, {Groff}, {Kasdin}, {McElwain}, {Singh}, {Uyama}, {Kuzuhara},
  {Akiyama}, {Grady}, {Hayashi}, {Knapp}, {Kwon}, {Oh}, {Wisniewski}, {Sitko},
  \& {Yang}}]{currie2017SSF}
{Currie}, T., {Guyon}, O., {Tamura}, M., {et~al.} 2017, \apjl, 836, L15,
  \dodoi{10.3847/2041-8213/836/1/L15}

\bibitem[{Dorrer \& Zuegel(2007)}]{Dorrer2007}
Dorrer, C., \& Zuegel, J.~D. 2007, J. Opt. Soc. Am. B, 24, 1268,
  \dodoi{10.1364/JOSAB.24.001268}

\bibitem[{{Floyd} \& {Steinberg}(1976)}]{FloydSteinberg1976}
{Floyd}, R.~W., \& {Steinberg}, L. 1976, Proc. Soc. Inf. Disp., 17, 75

\bibitem[{{Groff} {et~al.}(2016){Groff}, {Chilcote}, {Kasdin}, {Galvin},
  {Loomis}, {Carr}, {Brand t}, {Knapp}, {Limbach}, {Guyon}, {Jovanovic},
  {McElwain}, {Takato}, \& {Hayashi}}]{groff2016LTP}
{Groff}, T.~D., {Chilcote}, J., {Kasdin}, N.~J., {et~al.} 2016, in Society of
  Photo-Optical Instrumentation Engineers (SPIE) Conference Series, Vol. 9908,
  Ground-based and Airborne Instrumentation for Astronomy VI, 99080O,
  \dodoi{10.1117/12.2233447}

\bibitem[{{Guyon}(2003)}]{guyon2003PIA}
{Guyon}, O. 2003, \aap, 404, 379, \dodoi{10.1051/0004-6361:20030457}

\bibitem[{{Jovanovic} {et~al.}(2015){Jovanovic}, {Martinache}, {Guyon},
  {Clergeon}, {Singh}, {Kudo}, {Garrel}, {Newman}, {Doughty}, {Lozi}, {Males},
  {Minowa}, {Hayano}, {Takato}, {Morino}, {Kuhn}, {Serabyn}, {Norris},
  {Tuthill}, {Schworer}, {Stewart}, {Close}, {Huby}, {Perrin}, {Lacour},
  {Gauchet}, {Vievard}, {Murakami}, {Oshiyama}, {Baba}, {Matsuo}, {Nishikawa},
  {Tamura}, {Lai}, {Marchis}, {Duchene}, {Kotani}, \&
  {Woillez}}]{jovanovic2015SCE}
{Jovanovic}, N., {Martinache}, F., {Guyon}, O., {et~al.} 2015, \pasp, 127, 890,
  \dodoi{10.1086/682989}

\bibitem[{{Jovanovic} {et~al.}(2017{\natexlab{a}}){Jovanovic}, {Guyon},
  {Kotani}, {Kawahara}, {Hosokawa}, {Lozi}, {Males}, {Ireland}, {Tamura},
  {Mawet}, {Schwab}, {Norris}, {Leon-Saval}, {Betters}, \&
  {Tuthill}}]{jovanovic2017DPC}
{Jovanovic}, N., {Guyon}, O., {Kotani}, T., {et~al.} 2017{\natexlab{a}}, arXiv
  e-prints, arXiv:1712.07762.
\newblock \doarXiv{1712.07762}

\bibitem[{{Jovanovic} {et~al.}(2017{\natexlab{b}}){Jovanovic}, {Schwab},
  {Guyon}, {Lozi}, {Cvetojevic}, {Martinache}, {Leon-Saval}, {Norris}, {Gross},
  {Doughty}, {Currie}, \& {Takato}}]{jovanovic2017EIL}
{Jovanovic}, N., {Schwab}, C., {Guyon}, O., {et~al.} 2017{\natexlab{b}}, \aap,
  604, A122, \dodoi{10.1051/0004-6361/201630351}

\bibitem[{Jovanovic {et~al.}(2019)Jovanovic, Delorme, Bond, Cetre, Mawet,
  Echeverri, Wallace, Bartos, Lilley, Ragland, Ruane, Wizinowich, Chun, Wang,
  Wang, Fitzgerald, Pezzato, Matthews, Calvin, Millar-Blanchaer, Martin,
  Wetherell, Wang, Jacobson, Warmbier, Lockhart, Hall, Jensen-Clem, \&
  McEwen}]{jovanovic2019-KPI}
Jovanovic, N., Delorme, J.-R., Bond, C.~Z., {et~al.} 2019, in Techniques and
  Instrumentation for Detection of Exoplanets IX, ed. S.~B. Shaklan, Vol.
  11117, International Society for Optics and Photonics (SPIE),
  \dodoi{10.1117/12.2529330}

\bibitem[{Kasdin {et~al.}(2003)Kasdin, Vanderbei, Spergel, \&
  Littman}]{Kasdin2003}
Kasdin, N.~J., Vanderbei, R.~J., Spergel, D.~N., \& Littman, M.~G. 2003,
  Astrophys. J., 582, 1147

\bibitem[{{Macintosh} {et~al.}(2014){Macintosh}, {Graham}, {Ingraham},
  {Konopacky}, {Marois}, {Perrin}, {Poyneer}, {Bauman}, {Barman}, {Burrows},
  {Cardwell}, {Chilcote}, {De Rosa}, {Dillon}, {Doyon}, {Dunn}, {Erikson},
  {Fitzgerald}, {Gavel}, {Goodsell}, {Hartung}, {Hibon}, {Kalas}, {Larkin},
  {Maire}, {Marchis}, {Marley}, {McBride}, {Millar-Blanchaer}, {Morzinski},
  {Norton}, {Oppenheimer}, {Palmer}, {Patience}, {Pueyo}, {Rantakyro},
  {Sadakuni}, {Saddlemyer}, {Savransky}, {Serio}, {Soummer},
  {Sivaramakrishnan}, {Song}, {Thomas}, {Wallace}, {Wiktorowicz}, \&
  {Wolff}}]{macintosh2014GPI}
{Macintosh}, B., {Graham}, J.~R., {Ingraham}, P., {et~al.} 2014, Proceedings of
  the National Academy of Science, 111, 12661, \dodoi{10.1073/pnas.1304215111}

\bibitem[{{Males} {et~al.}(2018){Males}, {Close}, {Miller}, {Schatz},
  {Doelman}, {Lumbres}, {Snik}, {Rodack}, {Knight}, {Van Gorkom}, {Long},
  {Hedglen}, {Kautz}, {Jovanovic}, {Morzinski}, {Guyon}, {Douglas}, {Follette},
  {Lozi}, {Bohlman}, {Durney}, {Gasho}, {Hinz}, {Ireland}, {Jean}, {Keller},
  {Kenworthy}, {Mazin}, {Noenickx}, {Alfred}, {Perez}, {Sanchez}, {Sauve},
  {Weinberger}, \& {Conrad}}]{males2018MAG}
{Males}, J.~R., {Close}, L.~M., {Miller}, K., {et~al.} 2018, in Society of
  Photo-Optical Instrumentation Engineers (SPIE) Conference Series, Vol. 10703,
  Adaptive Optics Systems VI, 1070309, \dodoi{10.1117/12.2312992}

\bibitem[{{Martinez} {et~al.}(2009{\natexlab{a}}){Martinez}, {Dorrer}, {Aller
  Carpentier}, {Kasper}, {Boccaletti}, {Dohlen}, \&
  {Yaitskova}}]{Martinez2009a}
{Martinez}, P., {Dorrer}, C., {Aller Carpentier}, E., {et~al.}
  2009{\natexlab{a}}, Astron. Astrophys., 495, 363,
  \dodoi{10.1051/0004-6361:200810918}

\bibitem[{{Martinez} {et~al.}(2009{\natexlab{b}}){Martinez}, {Dorrer},
  {Kasper}, {Boccaletti}, \& {Dohlen}}]{Martinez2009b}
{Martinez}, P., {Dorrer}, C., {Kasper}, M., {Boccaletti}, A., \& {Dohlen}, K.
  2009{\natexlab{b}}, Astron. Astrophys., 500, 1281,
  \dodoi{10.1051/0004-6361/200911824}

\bibitem[{{Mawet} {et~al.}(2017){Mawet}, {Ruane}, {Xuan}, {Echeverri},
  {Klimovich}, {Randolph}, {Fucik}, {Wallace}, {Wang}, {Vasisht}, {Dekany},
  {Mennesson}, {Choquet}, {Delorme}, \& {Serabyn}}]{mawet2017OEH}
{Mawet}, D., {Ruane}, G., {Xuan}, W., {et~al.} 2017, \apj, 838, 92,
  \dodoi{10.3847/1538-4357/aa647f}

\bibitem[{Nisenson \& Papaliolios(2001)}]{Nisenson2001}
Nisenson, P., \& Papaliolios, C. 2001, Astrophys. J. Lett., 548, L201.
\newblock \url{http://stacks.iop.org/1538-4357/548/i=2/a=L201}

\bibitem[{{Sayson} {et~al.}(2019){Sayson}, {Ruane}, {Mawet}, {Jovanovic},
  {Calvin}, {Levraud}, {Roberson}, {Delorme}, {Echeverri}, {Klimovich}, \&
  {Xin}}]{Sayson2019}
{Sayson}, J.~L., {Ruane}, G., {Mawet}, D., {et~al.} 2019, Journal of
  Astronomical Telescopes, Instruments, and Systems, 5, 019004,
  \dodoi{10.1117/1.JATIS.5.1.019004}

\bibitem[{{Schwab} {et~al.}(2012){Schwab}, {Leon-Saval}, {Betters},
  {Bland-Hawthorn}, \& {Mahadevan}}]{schwab2012-SME}
{Schwab}, C., {Leon-Saval}, S.~G., {Betters}, C.~H., {Bland-Hawthorn}, J., \&
  {Mahadevan}, S. 2012, arXiv e-prints, arXiv:1212.4867.
\newblock \doarXiv{1212.4867}

\bibitem[{{Snellen} {et~al.}(2015){Snellen}, {de Kok}, {Birkby}, {Brandl},
  {Brogi}, {Keller}, {Kenworthy}, {Schwarz}, \& {Stuik}}]{snellen2015CHD}
{Snellen}, I., {de Kok}, R., {Birkby}, J.~L., {et~al.} 2015, \aap, 576, A59,
  \dodoi{10.1051/0004-6361/201425018}

\bibitem[{{Soummer} {et~al.}(2003){Soummer}, {Aime}, \&
  {Falloon}}]{Soummer2003_APLC}
{Soummer}, R., {Aime}, C., \& {Falloon}, P.~E. 2003, Astron. Astrophys., 397,
  1161, \dodoi{10.1051/0004-6361:20021573}

\bibitem[{{Sparks} \& {Ford}(2002)}]{Sparks2002}
{Sparks}, W.~B., \& {Ford}, H.~C. 2002, \apj, 578, 543, \dodoi{10.1086/342401}

\bibitem[{{Vigan} {et~al.}(2018){Vigan}, {Otten}, {Muslimov}, {Dohlen},
  {Philipps}, {Seemann}, {Beuzit}, {Dorn}, {Kasper}, {Mouillet}, {Baraffe}, \&
  {Reiners}}]{vigan2018BHS}
{Vigan}, A., {Otten}, G.~P.~P.~L., {Muslimov}, E., {et~al.} 2018, in Society of
  Photo-Optical Instrumentation Engineers (SPIE) Conference Series, Vol. 10702,
  Ground-based and Airborne Instrumentation for Astronomy VII, 1070236,
  \dodoi{10.1117/12.2313681}

\bibitem[{{Wang} {et~al.}(2017){Wang}, {Mawet}, {Ruane}, {Hu}, \&
  {Benneke}}]{wang2017OEH}
{Wang}, J., {Mawet}, D., {Ruane}, G., {Hu}, R., \& {Benneke}, B. 2017, \aj,
  153, 183, \dodoi{10.3847/1538-3881/aa6474}

\bibitem[{Watson {et~al.}(1991)Watson, Mills, Gaiser, \& Diner}]{Watson1991}
Watson, S.~M., Mills, J.~P., Gaiser, S.~L., \& Diner, D.~J. 1991, Appl. Opt.,
  30, 3253, \dodoi{10.1364/AO.30.003253}

\bibitem[{Woillez {et~al.}(2003)Woillez, Guerin, Perrin, Lai, Reynaud, Collin,
  Cretenet, Marlot, Pau, Reess, Ziegler, Berthod, \& Brient}]{woillez2003-IIS}
Woillez, J., Guerin, J., Perrin, G.~S., {et~al.} 2003, in Interferometry for
  Optical Astronomy II, ed. W.~A. Traub, Vol. 4838, International Society for
  Optics and Photonics (SPIE), 1341 -- 1343, \dodoi{10.1117/12.459362}

\bibitem[{{Zhang} {et~al.}(2018){Zhang}, {Ruane}, {Delorme}, {Mawet},
  {Jovanovic}, {Jewell}, {Shaklan}, \& {Wallace}}]{zhang2018CMA}
{Zhang}, M., {Ruane}, G., {Delorme}, J.-R., {et~al.} 2018, in Society of
  Photo-Optical Instrumentation Engineers (SPIE) Conference Series, Vol. 10698,
  Space Telescopes and Instrumentation 2018: Optical, Infrared, and Millimeter
  Wave, 106985X, \dodoi{10.1117/12.2312831}

\end{thebibliography}
\bibliographystyle{aasjournal}

\end{document}